\documentclass[journal]{IEEEtran} 

\usepackage{cite}
\usepackage{algorithm,algorithmicx}
\usepackage{algpseudocode}
\usepackage[small,bf]{caption}
\usepackage[cmex10]{amsmath}
\usepackage{amssymb,latexsym,epsfig,subfigure,epic,amscd,mathrsfs,euscript,eufrak,bm}
\usepackage{color}
\usepackage{booktabs}
\usepackage{array}

\usepackage{enumerate}
\usepackage{amsfonts}
\usepackage{amsopn}
\usepackage{graphicx}
\usepackage{url} 
\usepackage{empheq}
\usepackage{bbm}

\newtheorem{theorem}{\bf{Theorem}}
\newtheorem{remark}{\bf{Remark}}

\newtheorem{defn}{\bf{Definition}}

\DeclareMathOperator*{\Gtrless}{\gtrless}
\newcommand{\bs}[1]{ \ensuremath{ \boldsymbol{#1} }}

\begin{document}

\title{Fusion of Correlated Decisions Using Regular Vine Copulas}

\author{Shan~Zhang,
         Lakshmi~Narasimhan~Theagarajan,~\IEEEmembership{Member,~IEEE,}
         Sora~Choi 
	 and Pramod~K.~Varshney,~\IEEEmembership{Life Fellow,~IEEE}	
\thanks{Copyright (c) 2015 IEEE. Personal use of this material is permitted. However, permission to use this material for any other purposes must be obtained from the IEEE by sending a request to pubs-permissions@ieee.org.}
\thanks{S. Zhang and P. K. Varshney are with the Department
of EECS, Syracuse University, Syracuse, NY 13244, USA. Email: \{szhang60, varshney\}@syr.edu.}
\thanks{L. N. Theagarajan is with the Department of Electrical Engineering, Indian Institute of Technology Palakkad, Kozhippara, Kerala 678557, India. Email: lnt@iitpkd.ac.in.}
\thanks{S. Choi is with Delphi Electronics \& Safety, Kokomo, IN 46902, USA. Email: ssrchoi7@hotmail.com.}
\thanks{This work was supported by ARO under grant W911NF-14-1-0339 and AFOSR under grant FA9550-16-1-0077.}
}

\maketitle

\begin{abstract}
In this paper, we propose a regular vine copula based methodology for the fusion of correlated decisions. Regular vine copula is an extremely flexible and powerful graphical model to characterize complex dependence among multiple modalities. It can express a multivariate copula by using a cascade of bivariate copulas, the so-called pair copulas.  Assuming that local detectors are single threshold binary quantizers and taking complex dependence among sensor decisions into account, we design an optimal fusion rule using a regular vine copula under the Neyman-Pearson framework. In order to reduce the computational complexity resulting from the complex dependence, we propose an efficient and computationally light regular vine copula based optimal fusion algorithm. Numerical experiments are conducted to demonstrate the effectiveness of our approach.
\end{abstract}

\begin{IEEEkeywords}
Distributed detection, dependence modeling, regular vine copula, sensor fusion, decision fusion
\end{IEEEkeywords}

\section{Introduction}
\label{sec:I}
The problem of distributed detection has attracted significant attention over the past several decades. Distributed sensor networks consist of a large number of spatially distributed sensors that operate collaboratively to solve an inference problem. The dispersed sensors make noisy observations of a phenomenon of interest, and then transmit a compressed version of its data to the fusion center (FC) which fuses the received quantized data and produces a global decision. Distributed processing schemes in sensor networks have several advantages over centralized schemes, e.g., reduced communication cost, increased reliability and greater coverage of the network. In this paper, we study the problem of distributed detection with dependent sensor observations. In particular, we focus on a two-level parallel distributed detection system that consists of a number of local sensors and a FC. 
Each local sensor acquires its local observation of the phenomenon of interest and transmits a binary compressed version of the message to the FC, which detects the presence or absence of the target.

The problem of distributed detection with independent observations has been studied extensively \cite{tenney1981detection, chair1986optimal, hoballah1986neyman, viswanathan1997distributed, varshney2012distributed, chen2005optimality, niu2005distributed, niu2006distributed, chamberland2004asymptotic, yang2010distributed, fang2012optimal}. In \cite{tenney1981detection}, decision rules at the sensors were designed for the two-sensor distributed detection problem with conditionally independent observations. In \cite{chair1986optimal}, the optimal fusion rule given the local detectors was proposed. In \cite{hoballah1986neyman}, two problems were studied under the Neyman-Pearson framework for multiple sensors: design of local sensor decision rules given the fusion rule and design of the fusion rule given the local decision rules. In \cite{viswanathan1997distributed}, parallel and serial distributed detection systems were studied in some detail and local sensor decision rules were obtained under both Neyman-Pearson and Bayesian criteria. In \cite{varshney2012distributed}, the design of local sensor decision rules and the optimal fusion rule for the problem of distributed detection with binary local decisions for the parallel fusion system were discussed in great detail. Under Bayesian and Neyman-Pearson frameworks, it has been shown in \cite{varshney2012distributed} that the optimal sensor decision rule is the likelihood-ratio-based binary quantizer, and the optimal fusion statistic is a weighted sum of sensor decisions. In \cite{chen2005optimality}, the design of local decision rules in the presence of non ideal transmission channels between the local sensors and the FC was considered. In \cite{niu2005distributed, niu2006distributed}, distributed detection and decision fusion schemes were proposed for large random sensor networks. In \cite{chamberland2004asymptotic, yang2010distributed, fang2012optimal}, power constrained sensor networks for distributed detection systems were studied. In the aforementioned literature \cite{tenney1981detection, chair1986optimal, hoballah1986neyman, viswanathan1997distributed, varshney2012distributed, chen2005optimality, niu2005distributed, niu2006distributed, chamberland2004asymptotic, yang2010distributed, fang2012optimal}, sensor observations were assumed to be conditionally independent. However, for sensors deployed in practical distributed sensor networks, the observations are often dependent due to a variety of reasons such as sensing of the same phenomenon and dependent transmission channels. In this paper, we take the dependence among sensor observations into account and seek an optimal fusion rule for the detection of a random signal assuming that the transmission channels between the local sensors and the FC are ideal.

The problem of distributed detection with dependent observations has attracted some attention \cite{drakopoulos1991optimum, kam1992optimal, willett2000good, yan2001distributed, chamberland2006dense, khalid2012cooperative, kasasbeh2016hard, kasasbeh2017soft, veeravalli2012distributed}. In \cite{drakopoulos1991optimum, kam1992optimal}, optimal fusion rules for correlated binary decisions have been proposed. However, both approaches for correlated decision fusion require some prior information about the joint statistics of sensor observations or decisions. In \cite{willett2000good}, with correlated Gaussian observations, optimal Bayesian binary quantizers were designed for a two-sensor setting and the fusion rules $\mathnormal{\text{and}}$, $\mathnormal{\text{or}}$ and $\mathnormal{\text{xor}}$ were studied. However, it was found that the optimal fusion rule with dependent observations is much more complicated beyond the two-sensor setting. In \cite{yan2001distributed}, given a fixed fusion rule, optimum Neyman-Pearson distributed signal detection with correlated Gaussian noise was considered for multiple sensors. In \cite{chamberland2006dense}, the density of the network was studied for the problem of distributed detection with correlated Gaussian noise, where local sensors were assumed to be power constrained. In \cite{khalid2012cooperative, kasasbeh2016hard}, noisy correlated sensing channels were investigated for binary decision based distributed detection in multi-sensor systems where the majority rule was used at the FC to fuse local decisions. In \cite{drakopoulos1991optimum, kam1992optimal, willett2000good, yan2001distributed, chamberland2006dense, khalid2012cooperative, kasasbeh2016hard}, the local detectors were constrained to be binary quantizers. Moreover, the use of $\mathnormal{\text{and}}$, $\mathnormal{\text{or}}$, $\mathnormal{\text{xor}}$ and the majority rules was found to be far from optimal when the dependence structure is complex. In \cite{kasasbeh2017soft}, noisy correlated sensing channels were studied for multi-bit decision based distributed detection and a likelihood ratio test was used to generate the global decision at the FC. In a recent survey paper \cite{veeravalli2012distributed}, additional work on distributed detection problem with dependent observations has been discussed. In the presence of dependence among sensors, the computational complexity of the distributed detection problem increases significantly. It has been shown that the distributed detection problem with dependent observations cannot be solved using a polynomial time algorithm \cite{tsitsiklis1985complexity}. Therefore, the design of optimal local decision rules may not be possible due to computational intractability resulting from the dependence among sensor observations. Thus, in this paper, we assume that local detectors are single threshold binary quantizers and derive an optimal fusion rule under the Neyman-Pearson framework.

In the existing work on distributed detection with dependent observations \cite{drakopoulos1991optimum, kam1992optimal, willett2000good, yan2001distributed, chamberland2006dense, khalid2012cooperative, kasasbeh2016hard, kasasbeh2017soft}, prior information of the joint statistics of sensor observations or decisions was assumed to be given. The fusion rule for multiple sensors requires a complete knowledge of the form and structure of the joint distribution of sensor observations. Generally, the joint statistics of sensor observations is not available \textit{a priori}. Moreover, the dependence structure of multivariate sensors can be quite complex and nonlinear. Simple dependence modeling through methods such as the use of multivariate normal model, is very limited and inadequate to characterize complex dependence among multiple sensors. 

Copula-based dependence modeling \cite{nelsen2013introduction} is a flexible parametric characterization of the joint distribution of sensor observations. Using copula-based dependence modeling, approximate joint distribution functions can be constructed from arbitrary marginal distributions. Moreover, it allows separation of modeling univariate marginals from modeling the multivariate (dependence) structure. It has been shown that copula-based fusion of multiple sensing observations can significantly improve the performance of inference problems \cite{iyengar2011parametric, he2012fusing, sundaresan2011location, iyengar2011biometric, he2015social}. However, the class of known multivariate copulas required for the fusion of observations from more than two sensors is limited. Gaussian copulas perform poorly on data with heavy tails. Student-$t$ copulas allow for symmetric tail dependence, but they have only a single parameter to capture tail dependence among all the variables. While standard Archimedean multivariate copulas can characterize asymmetric tail dependence, they are quite limited as they are characterized by only a single parameter. This shows that there is a growing need for more flexible copulas especially for modeling high-dimensional dependence structures. Regular vine (R-Vine) copulas \cite{bedford2001probability, bedford2002vines, aas2009pair} are graphical models constructed to overcome the limitations of the existing standard multivariate copulas. They are hierarchical in nature since they can express a multivariate copula by using a cascade of bivariate copulas, the so-called pair copulas. Canonical vines or C-Vines and Drawable vines or D-Vines, two types of R-Vines, have been analyzed in \cite{aas2009pair}.
In \cite{subramanian2011fusion}, D-Vine copula based fusion of dependent signals was proposed for the detection problem. Different from \cite{subramanian2011fusion}, here we propose a more flexible optimal R-Vine copula based fusion rule with binary quantizers for the parallel distributed detection system. 

In \cite{sundaresan2011copula}, a copula-based fusion methodology for correlated decisions was proposed in the Neyman-Pearson framework for the problem of distributed detection. However, it mainly focused on the fusion for the two-sensor case. Compared to \cite{sundaresan2011copula}, in this paper, we consider a more general problem than \cite{sundaresan2011copula}. We propose a novel and powerful fusion methodology for the fusion of dependent decisions, R-Vine copula based fusion, for more flexible modeling of complex dependency especially for larger number of sensors. Note that our R-Vine copula based fusion methodology is a data-driven approach so that sufficient amount of data is required for modeling purposes. We summarize our contributions as follows. 
\begin{itemize}
\item We take complex dependence into account for the fusion of dependent decisions and use regular vine copula to model dependence that enables the fusion of more than two decisions. 
\item We propose an optimal regular vine copula based fusion rule at the FC assuming that the local detectors are single threshold binary quantizers. Furthermore, we propose an efficient fusion algorithm which significantly reduces the computational complexity resulting from the dependence that exists among sensor observations. 
\item We show the superiority of our proposed regular vine copula based fusion methodology via a number of illustrative examples.  
\end{itemize}

The rest of the paper is organized as follows. In Section\, \ref{sec: CT}, we provide a brief introduction to copula theory including Sklar's Theorem and Vine copulas. In Section\,\ref{sec: PF}, we introduce the parallel distributed detection system, and state the distributed detection problem. In Section\,\ref{sec: R-Vine}, we propose an optimal regular vine copula based fusion rule. Also, to reduce the computational complexity, we further propose an efficient fusion algorithm.  In Section\,\ref{sec:nr}, we demonstrate the effectiveness of the proposed R-Vine copula based fusion algorithm through numerical examples. Finally, in Section\ \ref{sec: conclusion} we summarize our work and discuss future research directions.



\section{Copula Theory Background}
\label{sec: CT}
Dependence modeling with copulas provides a flexible and powerful approach for modeling continuous multivariate distributions since it separates modeling univariate marginals from modeling the multivariate (dependence) structure. A standard multivariate copula, specified independently from marginals, is a multivariate distribution with uniform marginal distributions. The unique correspondence between a standard multivariate copula and any multivariate distribution is stated in Sklar's theorem \cite{nelsen2013introduction} which is a fundamental theorem that forms the basis of copula theory. Standard multivariate copulas lack the flexibility to model complex dependencies due to factors such as limited number of parameters to characterize dependence. Vine copula methodology has been developed for more flexible modeling of complex dependencies in larger dimensions. In the following, we first give the theoretical background of standard multivariate copulas, and then introduce the regular vine copula which we will use in this paper.

\subsection{Standard Multivariate Copulas}
\label{subsec:st}
\begin{theorem}[Sklar's Theorem]
The joint distribution function $F$ of random variables $x_1,\ldots,x_d$ with continuous marginal distribution functions $F_1,\ldots,F_d$ can be cast as
\begin{equation}
\label{CopEq1}
F(x_1,x_2,\ldots,x_d) = C(F_1(x_1),F_2(x_2),\ldots,F_d(x_d)),
\end{equation}
where $C$ is a unique standard $d$-dimensional copula. Conversely, given a copula $C$ and univariate Cumulative Distribution Functions (CDFs) $F_1,\ldots,F_d$, $F$ in \eqref{CopEq1} is a valid multivariate CDF with marginals $F_1,\ldots,F_d$.
\end{theorem}

For absolutely continuous distributions $F$ and $F_1,\ldots,F_d$, the joint Probability Density Function (PDF) of random variables $x_1,\ldots,x_d$ can be obtained by differentiating both sides of \eqref{CopEq1}:
\begin{equation}
\label{CopEq2}
f(x_1,\ldots,x_d) = \Big(\prod_{m=1}^{d}f_m(x_m)\Big)c(F_1(x_1),\ldots,F_d(x_d)),
\end{equation}
where $f_1, \ldots, f_d$ are the marginal densities and $c$ is referred to as the density of standard multivariate copula $C$ that is given by
\begin{equation}
\label{CopDens}
c(\mathbf{u}) = \frac{\partial^L(C(u_1,\ldots,u_d))}{\partial u_1,\ldots,\partial u_d},
\end{equation}
where $u_m=F_m(x_m)$ and $\mathbf{u} = [u_1,\dots,u_d]$.

Thus, given specified univariate marginal distributions $F_1,\ldots,F_d$ and copula model $C$, the joint distribution function $F$ can be constructed by
\begin{equation}
F(F_1^{-1}(u_1), F_2^{-1}(u_2), \ldots, F_d^{-1}(u_d)) = C(u_1,u_2, \ldots, u_d),
\end{equation}
where $u_m=F_m(x_m)$ and $F_m^{-1}(u_m)$ are the inverse distribution functions of the marginals, $m = 1, 2, \ldots, d$.

Note that $C(\cdot)$ is a valid CDF and $c(\cdot)$ is a valid PDF for uniformly distributed random variables $u_m$, $m = 1, 2, \ldots, d$. Since the random variable $u_m$ represents the CDF of $x_m$, the CDF of $u_m$ naturally follows a uniform distribution over $[ 0,1]$.

Since different copula functions may model different types of dependence, selection of copula functions to characterize joint statistics of random variables is a key problem. Various families of standard multivariate copula functions are described in \cite{nelsen2013introduction}, of which the elliptical and Archimedean copulas (see Appendix\,\ref{appendix:copula} for details) are widely used.
Moreover, \textit{dependence parameter} denoted by $\bs{\phi}$, contained in a copula function, is used to characterize the amount of dependence among $d$ random variables. Typically, $\bs{\phi}$ is unknown a \textit{priori} and needs to be estimated, e.g., using Maximum Likelihood Estimation (MLE) or Kendall's $\tau$~\cite{he2015heterogeneous}. Note that in general, $\bs{\phi}$ may be a scalar, a vector or a matrix.

\subsection{Vine Copulas}
\label{subsec:vine}
Regular vine (R-Vine) copulas are extremely flexible in modeling multivariate dependence especially in high dimensions \cite{dissmann2013selecting, mari2001correlation}, where a set of bivariate copulas are used at different hierarchical levels. Regular vine, introduced by Bedford and Cooke in \cite{bedford2001probability, bedford2002vines}, is a graphical model that is convenient to describe the dependence structure of random variables. 
A regular vine is defined as follows.
\begin{defn}[R-Vine] $\mathcal{V}=(T_1,\dots,T_{d-1})$ is a regular vine on $d$ elements if the following conditions are satisfied.
\begin{enumerate}
\item $T_1$ is a tree with nodes $N_1=\{1,\dots,d\}$ and a set of $d - 1$ edges denoted as $E_1$.
\item For $i=2,\dots,d-1$, $T_i$ is a tree with nodes $N_i=E_{i-1}$ and edge set $E_i$.
\item For $i=2, \dots, d-1$ and $\{a,b\} \in E_i$ with $a=\{a_1,a_2\}$ and $b=\{b_1,b_2\}$, $|(a \cap b)|=1$ (proximity condition) holds, where $|\cdot|$ denotes the cardinality of a set.
\end{enumerate}
\end{defn}
A $d$-dimensional vine consists of $d(d-1)/2$ edges in total. The proximity condition implies that two edges in tree $T_i$ are connected in tree $T_{i + 1}$ if the two edges share a common node in tree $T_i$. 

\begin{figure}
\centering
\includegraphics[height=2.3in,width=!]{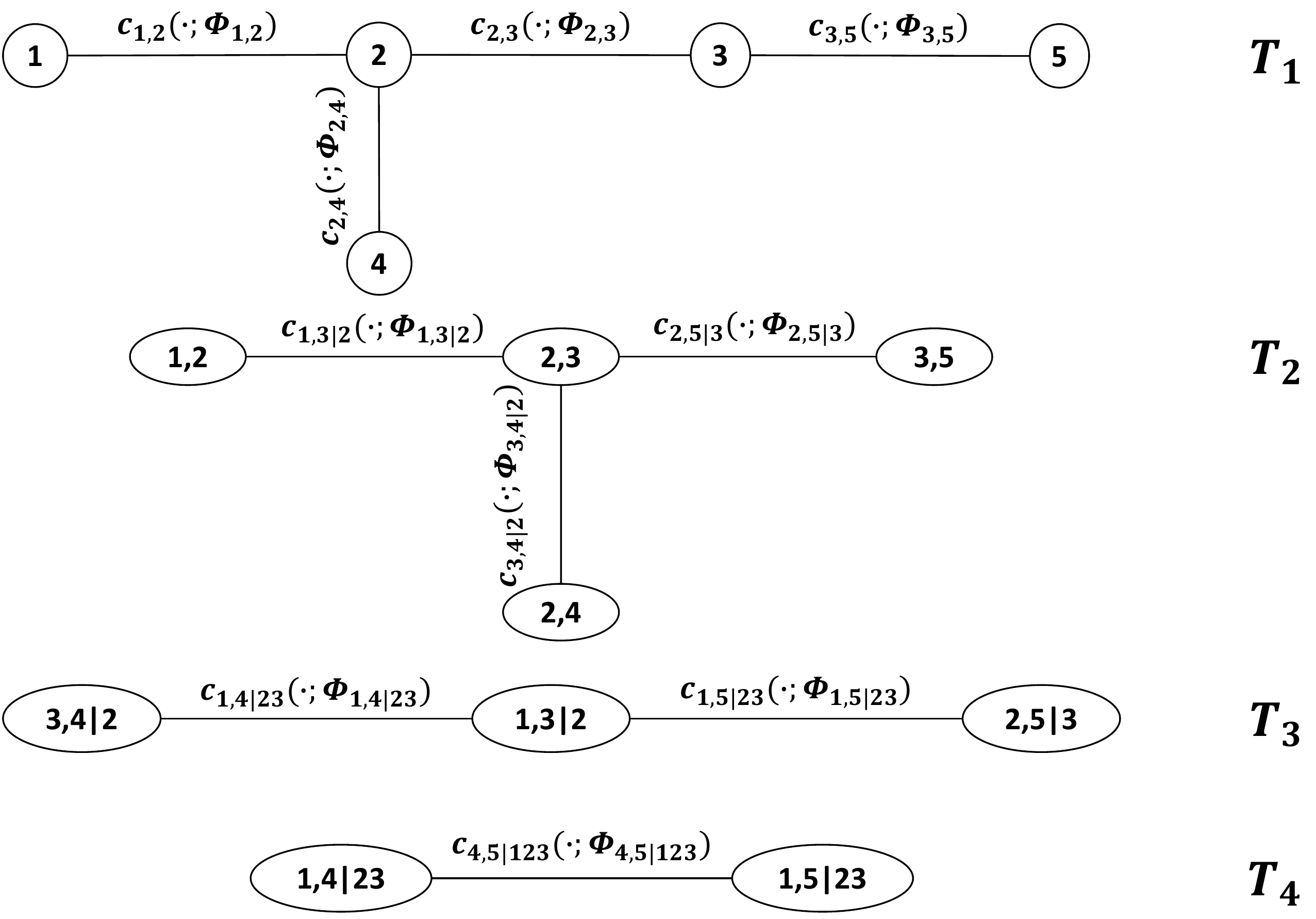}
\caption{An example R-Vine for five variables.}
\label{fig: example_RVine}
\end{figure}

R-Vine copula is obtained by specifying bivariate copulas, the so-called pair-copula, on each of the edges. Before introducing R-Vine copula, some sets associated with its edges need to be defined. The complete union $U_e$ of an edge $e = \{ a, b \} \in E_i, a, b \in N_i$ is defined as $U_e = \{ m \in N_1 \, | \, \exists e_j \in E_j, j = 1, 2, \ldots, i - 1, \text{such that} \, m \in e_1 \in \ldots e_{i-1} \in e \}$. The conditioning set of the edge $e = \{ a, b \}$ is $D_e = U_a \cap U_b$ and the conditioned sets of the edge $e = \{ a, b \}$ are $\mathfrak{C}_{e, a} = U_a \backslash D_e$ and $\mathfrak{C}_{e, b} = U_b \backslash D_e$; see an illustrative example in Fig.\,\ref{fig: example_RVine}. A regular vine copula is defined as follows, more details are provided later in Section\, \ref{sec: R-Vine}.

\begin{defn}[R-Vine Copula]
\label{def:R-VineC}
$(\mathbf{F}, \mathcal{V}, \mathbf{B})$ is called a R-Vine copula if
\begin{enumerate}
\item $\mathbf{F} = [F_1, F_2, \ldots, F_d]^T \in [0, 1]^d$ is a vector with uniform marginals.
\item $\mathcal{V}$ is a $d$-dimensional regular vine.
\item $\mathbf{B} = \{ C_{\mathfrak{C}_{e, a}, \mathfrak{C}_{e, b} | D_e} \, | \, e \in E_i, i = 1, 2, \ldots, d - 1 \}$ is a set of bivariate copulas. 
\end{enumerate}
\end{defn}

The joint density of a random vector $\mathbf{x} = [x_1, x_2, \ldots, x_d]^T$ is given by
\begin{align}
\label{eq:densityvine}
&f_{1,\dots,d}(\mathbf{x})=\prod_{m=1}^d f_m(x_m)\prod_{i=1}^{d-1}\prod_{e\in E_i} \times  \\ \nonumber
&c_{\mathfrak{C}_{e, a}, \mathfrak{C}_{e, b} | D_e}(F_{\mathfrak{C}_{e, a} | D_e}(x_{\mathfrak{C}_{e, a}} | \mathbf{x}_{D_e}),F_{\mathfrak{C}_{e, b} | D_e}(x_{\mathfrak{C}_{e, b}} | \mathbf{x}_{D_e})), \nonumber
\end{align}
where $e=\{a, b\}$, $\mathbf{x}_{D_e}=\{x_j | j \in D_e\}$, $f_m$ is the marginal density function of variable $x_m$, $m = 1, \dots, d$.  The conditional distribution $F_{\mathfrak{C}_{e, a} | D_e}(x_{\mathfrak{C}_{e, a}} | \mathbf{x}_{D_e})$ is obtained by the following equation \cite{joe1996families}.
\begin{align}
\label{eq:FF}
&F_{\mathfrak{C}_{e, a} | D_e}(x_{\mathfrak{C}_{e, a}} | \mathbf{x}_{D_e}) = \\ \nonumber
&\frac{\partial C_{\mathfrak{C}_{a, a_1}\!, \mathfrak{C}_{a, a_2}\! |D_a}\!\!\left(\!F_{\mathfrak{C}_{a, a_1}\! | D_a}\!(\!x_{\mathfrak{C}_{a, a_1}} \!| \mathbf{x}_{D_a}\!)\!, F_{\mathfrak{C}_{a, a_2}\! | D_a}\!(\!x_{\mathfrak{C}_{a, a_2}} | \mathbf{x}_{D_a}\!)\!\right)}{\partial F_{\mathfrak{C}_{a, a_2}\! | D_a}\!(\!x_{\mathfrak{C}_{a, a_2}}\! | \mathbf{x}_{D_a}\!)}, \nonumber
\end{align}
where $e = \{a, b\} \in E_i$, $a = \{a_1, a_2\}$ and $b = \{b_1, b_2\}$ are the edges that connect $\mathfrak{C}_{e, a}$ and $\mathfrak{C}_{e, b}$ given the conditioning variables $D_e$. Similarly, we can obtain $F_{\mathfrak{C}_{e, b} | D_e}(x_{\mathfrak{C}_{e, b}} | \mathbf{x}_{D_e})$.

As an example, a 5-dimensional R-Vine copula is shown in Fig.\,\ref{fig: example_RVine}. The R-Vine has four trees $T_i$ and the tree $T_i$ has nodes $N_i = 6 - i$ and edges $E_i = 5 - i$, where $i = 1, 2, 3, 4$. Each edge is associated with a bivariate copula density $c$ and its corresponding parameters $\bs{\phi}$ used to model  dependence between two variables. Moreover, at each edge $e = \{ a, b \} \in E_i$, the term $\mathfrak{C}_{e, a}$ and $\mathfrak{C}_{e, b}$ are separated by a comma and given to the left of the ``$|$" sign, while $D_e$ appears on the right.  In the first tree $T_1$, the dependences of the four pairs of variables $(1, 2), (2, 3), (2, 4), (3, 5)$ are modeled using four bivariate copulas, $c_{1,2}(\cdot ; \bs{\phi}_{1,2})$, $c_{2,3}(\cdot ; \bs{\phi}_{2,3})$, $c_{2,4}(\cdot ; \bs{\phi}_{2,4})$ and $c_{3,5}(\cdot ; \bs{\phi}_{3,5})$. In the second tree $T_2$, three conditional dependencies are modeled. The pair $(1, 3 | 2)$ using bivariate copula density $c_{1, 3 | 2}(\cdot ; \bs{\phi}_{1, 3 | 2})$ characterizes the dependence between the first and third variables given the second variable. Also, the pair $(3, 4 | 2)$ using bivariate copula density $c_{3, 4 | 2}(\cdot ; \bs{\phi}_{3, 4 | 2})$ characterizes the dependence between the third and fourth variables given the second variable. Similarly, we can obtain the bivariate copula density for the pair $(2, 5 | 3)$. In the third tree $T_3$, the dependence of the first and fourth variables given the second and third variables is modeled using bivariate copula density $c_{1, 4 | 23}(\cdot ; \bs{\phi}_{1, 4 | 23})$. Also, we can obtain the bivariate copula density for the pair $(1, 5 | 2 3)$. In the fourth tree $T_4$, the bivariate copula density $c_{4, 5 | 123}(\cdot ; \bs{\phi}_{4, 5 | 123})$ captures the dependence between the fourth and fifth variables given the first, second and third variables.

For the 5-dimensional case, using \eqref{eq:densityvine}, the joint PDF of $\mathbf z = [z_1, z_2, z_3, z_4, z_5]$ can be expressed as
\begin{equation*}
\begin{aligned}
&f(z_1, z_2, z_3, z_4, z_5) = \left [ \prod_{l=1}^5 f(z_{l}) \right] \cdot c_{1,2}\left (F(z_1), F(z_2); \bs{\phi}_{1,2} \right) \\
&\cdot c_{2,3}\left (F(z_2), F(z_3); \bs{\phi}_{2,3} \right) \cdot c_{2,4}\left (F(z_2), F(z_4); \bs{\phi}_{2,4} \right) \\
&\cdot c_{3,5}\left (F(z_2), F(z_3)\cdot ; \bs{\phi}_{3,5} \right) \\
&\cdot c_{1,3 | 2}\left (F(z_1 | z_2), F(z_3 | z_2); \bs{\phi}_{1,3 | 2} \right)\\
&\cdot c_{3,4 | 2}\left (F(z_3 | z_2), F(z_4 | z_2); \bs{\phi}_{3,4 | 2} \right) \\
&\cdot c_{2,5 | 3}\left (F(z_2 | z_3), F(z_5 | z_3); \bs{\phi}_{2,5 | 3} \right) \\
&\cdot c_{1,4 | 23}\left (F(z_1 | z_2 z_3), F(z_4 | z_2z_3); \bs{\phi}_{1,4 | 23} \right)
\end{aligned}
\end{equation*}

\begin{equation*}
\begin{aligned}
&\cdot c_{1,5 | 23}\left (F(z_1 | z_2 z_3), F(z_5 | z_2z_3); \bs{\phi}_{1,5 | 23} \right) \\
&\cdot c_{4,5 | 123}\left (F(z_4 | z_1z_2 z_3), F(z_5 | z_1z_2z_3); \bs{\phi}_{4,5 | 123} \right).
\end{aligned}
\end{equation*}

\subsection{Array representation of R-Vine}
Generally, it is quite expensive to store the nested set of trees and also not convenient to describe inference algorithms. In \cite{morales2010bayesian}, a lower triangular array was proposed to store an R-Vine. 

\begin{defn}[R-Vine Array] 
\label{def:R-VArr}
A lower triangular array $M = (m_{i,j})_{i, j = 1, 2, \ldots, d}$ is called an R-Vine array if for $i = 1, \ldots, d-1$ and for all $k = i+1, \ldots, d-1$, there is a $j$ in $i+1,\ldots, d-1$ with $(m_{k,i}, \{m_{k+1,i}, \ldots, m_{d,i}\}) \in B_M(j)$ or $\in \tilde{B}_M(j)$, 
where $B_M(j):= \{(m_{j,j}, D) | k=j+1, \ldots, d\}$ with $D = \{m_{k,j}, \ldots, m_{d,j} \}$ and $\tilde{B}_M(j):=  \{(m_{k,j}, \tilde{D}) | k=j+1, \ldots, d \}$ with $\tilde{D} = \{m_{j,j}\} \cup \{m_{k+1,j}, \ldots, m_{d,j}\}$.
\end{defn}

For the R-Vine copula example  in Fig.\,\ref{fig: example_RVine}, the R-Vine matrix $\mathbf M^*$ is given as
\begin{equation*}
\begin{bmatrix}
 5&  &  &  & \\ 
 4&  4&  &  & \\ 
 1&  1&  1&  & \\ 
 2&  3&  3& 3 & \\ 
 3&  2&  2&  2& 2 
\end{bmatrix},
\end{equation*}
where the first column represents the dependence of four pairs of variables, $(5,4 | 123)$, $(5, 1 | 23)$, $(5, 2 | 3)$ and $(5, 3)$. Going through all columns, we can see that the matrix $\mathbf M^*$ codes all information needed to represent
the R-vine copula in Fig.\,\ref{fig: example_RVine}.

 An R-Vine array has the following two properties:
\begin{itemize}
\item $\{m_{i,i}, \ldots, m_{d,i}\} \subset \{m_{j,j}, \ldots, m_{d,i}\}$ for $1 \leq j < i \leq d$,
\item $ m_{i,i} \notin \{m_{i+1,i+1},\ldots, m_{d,i+1}\}$ for $i = 1, \ldots, d-1$,
\end{itemize}
where the first property states that every column in the left contains all the entries that a column in the right contains, and the second property guarantees that there is a new entry on the diagonal in every column.

Given an R-Vine array $M=(m_{i,j})_{i,j=1, \ldots, d}$, the R-Vine distribution density \cite{dissmann2013selecting} is
\begin{align}
\label{Eq:RDistrDen}
&f_{1, \ldots, d} = \prod_{j=1}^d f_j \prod_{k=d-1}^{1}\prod_{i=d}^{k+1}c_{m_{k,k},m_{i,k} | m_{i+1,k}, \ldots, m_{d,k}} \\ \nonumber
&\left (F_{m_{k,k} | m_{i+1,k}, \ldots, m_{d,k}}, F_{m_{i,k} | m_{i+1,k}, \ldots, m_{d,k}}\right ). \nonumber
\end{align}

For notational simplicity, we have removed the arguments of all the functions in Equation \eqref{Eq:RDistrDen}.

%

\section{Problem Statement}
\label{sec: PF}
Consider a distributed detection problem, where a random phenomenon is monitored by $L$ sensors. 
A binary hypothesis testing problem is studied, where $H_1$ denotes the presence of the random phenomenon and $H_0$ denotes the absence of the phenomenon. The sensors make a set of observations at time instant $n$, $\mathbf{z}_{n} = [z_{1n}, z_{2n}, \ldots, z_{Ln}], n = 1, 2, \ldots, N$. We assume that the sensor observations are dependent across sensors. Moreover,
we further assume that the sensor observations are continuous random variables that are conditionally independent and identically distributed (i.i.d.) over time. Let $f(z_{ln} | H_1)$ and $f(z_{ln} | H_0)$ be the PDFs of the observation at the $l$th sensor and $n$th time instant under $H_1$ and $H_0$ hypotheses, respectively. No knowledge about the joint distribution of the sensor observations is available \textit{a priori}. Instead of transmitting noisy raw observations, local binary sensor decisions $u_{ln}$ are sent to the FC by using a binary quantizer which is defined as 
\begin{equation}
\label{eq:bq}
u_{ln} =
  \begin{cases}
    0       & \quad -\infty < z_{ln} < \tau_{l}\\
    1  & \quad \tau_{l} \leq z_{ln} < +\infty\\
    \end{cases},
\end{equation}
where $\tau_{l}$ is the quantizer threshold at the $l$th sensor. At the FC, local binary decisions are combined to obtain a global decision. 

Under the Neyman-Pearson criterion, the design problem for the parallel distributed detection system consists of deriving individual sensor thresholds $\tau_{l}$ to form sensor decisions and the optimal fusion rule that fuses local sensor decisions to obtain the global decision. The sensor thresholds $\tau_{l}, l = 1, 2, \ldots, L$ are obtained by maximizing the local probability of detection subject to a constraint on the local probability of false alarm. Note that these sensor thresholds are not necessarily optimal in the global sense. The design of the optimal fusion rule for multiple sensors is discussed next. 


Since sensor decisions are independent over time, the optimal test statistic \cite{varshney1997distributed} is given as
\begin{equation}
\label{eq:t}
\Lambda(\mathbf{u}) = \frac{\prod_{n=1}^{N}P(u_{1n}, u_{2n}, \ldots, u_{Ln}|H_1)}{\prod_{n=1}^{N}P(u_{1n}, u_{2n}, \ldots, u_{Ln}|H_0)},
\end{equation}
where $P(u_{1n}, u_{2n}, \ldots, u_{Ln}|H_k)$ is the joint probability mass function (PMF) of the sensor decisions at the $n$th time instant under $k$th hypothesis, $k = 0, 1$. We define $S = \{u_{1n}u_{2n} \ldots u_{Ln} | u_{ln} \in \{ 0, 1 \}, l = 1, 2, \ldots, L \}$ as the set of all permutations that specify $L$-sensor decisions at time instant $n$. There are a total of $2^L$ permutations for $L$ sensors. For a three-sensor problem, $S = \left\{ \{000\}, \{001\}, \{010\}, \{011\}, \{100\}, \{101\}, \{110\}, \{111\} \right\}$. Let
\begin{equation}
\begin{aligned}
&P(u_{1n}, u_{2n}, \ldots, u_{Ln}|H_1) = P_s, ~\text{and} \\
&P(u_{1n}, u_{2n}, \ldots, u_{Ln}|H_0) = Q_s,
\end{aligned}
\end{equation}
where $s \in S$. $P_s$ and $Q_s, s \in S$ are required while computing the test statistic at the FC. For a three-sensor problem, the set of probabilities  $P_{000}$, $P_{001}$, $P_{010}$, $\ldots$, $P_{111}$ and $Q_{000}$, $Q_{001}$, $Q_{010}$, $\ldots$, $Q_{111}$ that characterize the joint PMFs of sensor decisions $u_{1n}$, $u_{2n}$ and $u_{3n}$ under hypotheses $H_1$ and $H_0$, respectively, are needed. By integrating the joint PDFs of the sensor observations under both hypotheses, these probabilities can be obtained with the quantizer threshold $\tau_{l}$, $l = 1, 2, 3$. For example, 
\begin{equation}
\begin{aligned}
\label{eq:inte}
&P_{000} = \int_{z_{1} = - \infty}^{\tau_1}\int_{z_{2} = - \infty}^{\tau_2}\int_{z_{3} = - \infty}^{\tau_3}f(z_{1}, z_{2}, z_{3} | H_1)d{z_{1}}d{z_{2}}d{z_{3}}, \\
&P_{010} = \int_{z_{1} = - \infty}^{\tau_1}\int_{\tau_2}^{z_{2} = + \infty}\int_{z_{3} = - \infty}^{\tau_3} f(z_{1}, z_{2}, z_{3} | H_1)d{z_{1}}d{z_{2}}d{z_{3}},
\end{aligned}
\end{equation}
where for the simplification of notation, we omit the time index $n$ in the example.

However, due to existing complex and nonlinear dependence, the joint PDFs of sensor observations under both hypotheses are not known. Before determining the joint PMFs of sensor decisions, we first need to obtain the joint PDFs of sensor observations given only the knowledge of marginal PDFs of the sensor observations and the marginal PMFs of sensor decisions. Typically in many applications, we do not have any prior information related to the phenomenon of interest. Therefore, we may also need to determine the marginals of sensor observations. 

The dependence across sensors can be quite complicated and nonlinear. 
Assuming conditional independence among multiple sensors may result in substantial performance degradation. To design the optimal fusion rule, we propose a copula based fusion methodology to characterize the existing dependence and determine the joint PDFs of sensor observations. 
Due to the limitations of the class of standard multivariate copulas and complex dependence that generally exists among multiple sensors, more flexible dependence modeling approaches are needed to obtain the joint PDFs of sensor measurements. R-Vine copula based dependence modeling provides us a solution. It can express a multivariate copula using a cascade of bivariate copulas embedded in a tree structure that is shown to be more flexible and powerful to model the complex dependence. Note that learning of the joint distribution requires raw sensor observations. It can be done offline. Here, in our paper, we assume that the joint statistics of the sensors does not change over time. After measurement collection, raw measurements are sent to the FC. The FC uses these analog measurements to learn the joint statistics of the sensors. After that, only binary decisions are sent to the FC.

Taking the above considerations into account, in the following, we develop a novel and powerful R-Vine copula based fusion methodology for distributed detection. We will propose the optimal test statistic for the parallel distributed detection system and derive its asymptotic statistic. Furthermore, at the end, via simulations, we will show its power and flexibility to capture complex dependence and improve detection performance. Note that our proposed distributed detection system consists of three approximations. First, we constrain the local detectors to be binary quantizers. Second, we find the local thresholds using the Neyman-Pearson formulation at the sensors. These sensor thresholds are optimal at the local level but are not necessarily optimal for the global system. Third, we use regular vine copula based approach to approximate the joint PDF of sensor observations.

\section{R-Vine Copula Based Fusion of Multiple Correlated Decisions}
\label{sec: R-Vine}
\subsection{Optimal Test Statistic}
The optimal test statistic for $L$ sensors is characterized in \eqref{eq:t}. The joint PMF of $u_{ln}$, $l = 1, 2, \ldots, L$, at time $n$, $n = 1, 2, \ldots, N$ under $H_1$ and $H_0$, respectively, is given as: 
\begin{equation}
\label{eq:PQ}
\begin{aligned}
&P(u_{1n}, u_{2n}, \ldots, u_{Ln}|H_1) = \underset{s \in S}{\prod}P_{s}^{\prod_{l = 1}^{L}x_{ln}},  \\
&P(u_{1n}, u_{2n}, \ldots, u_{Ln}|H_0) = \underset{s \in S}{\prod}Q_{s}^{\prod_{l = 1}^{L}x_{ln}},
\end{aligned}
\end{equation}
where $s_l$ indicates the $l$th element of $s$, and $x_{ln} = u_{ln}$ if $s_{l} = 1$, otherwise, $x_{ln} = 1 - u_{ln}$ for $s \in S$. For example, see \eqref{eq:p} and \eqref{eq:q}, which are special cases of \eqref{eq:PQ} for $L = 3$.

Substituting \eqref{eq:PQ} in \eqref{eq:t} and taking log on both sides, the log test statistic is given by
\begin{align}
\label{eq:logT}
&\text{log}\Lambda(\mathbf{u}) = \!\!\!\sum_{\{ i_1n \} \in \mathnormal{I}_1 }\!\!\!A_{\mathbf{u}^1}\!\!\sum_{n = 1}^{N}\mathbf{u}^1\,\, + \!\!\!\!\!\sum_{\{i_1n, i_2n\} \in \mathnormal{I}_2 }\!\!\!\!A_{\mathbf{u}^2}\sum_{n = 1}^{N}\mathbf{u}^2 +\ldots + \\ \nonumber
&\sum_{\{i_1n, i_2n, \ldots, i_tn\} \in \mathnormal{I}_t }\!\!\!\!\!\!\!\!A_{\mathbf{u}^t}\sum_{n = 1}^{N}\mathbf{u}^t + \ldots + \!\!\!\!\!\!\!\!\!\!\sum_{\{i_1n, i_2n, \ldots, i_Ln\} \in \mathnormal{I}_L }\!\!\!\!\!\!\!\!\!A_{\mathbf{u}^L}\sum_{n = 1}^{N}\mathbf{u}^L \nonumber
\end{align}
where $\mathnormal{I} = \{ ln | u_{ln}\in \{0, 1\}, l = 1, 2, \ldots, L, n = 1, 2, \ldots, N \}$, $\mathnormal{I}_i$ is a subset of $\mathnormal{I}$ and the cardinality of the set $\mathnormal{I}_i$ is $i$, namely, $\left | \mathnormal{I}_{i} \right | = i$. Moreover, $\mathbf{u}^t = \{u_{i_1n} u_{i_2n}\ldots u_{i_tn}\}$, $t \in [1, 2, \ldots, L]$ and its weight is given as $A_{\mathbf{u}^t} = \text{log}\frac{\prod_{0 \leq k \leq t}\mathcal{P}_{\tilde{\mathnormal{I}}_{tk}^e}^{(-1)^{t}} \prod_{0 \leq k \leq t}\mathcal{Q}_{\tilde{\mathnormal{I}}_{tk}^o}^{(-1)^{t}}}{\prod_{0 \leq k \leq t}\mathcal{Q}_{\tilde{\mathnormal{I}}_{tk}^e}^{(-1)^{t}} \prod_{0 \leq k \leq t}\mathcal{P}_{\tilde{\mathnormal{I}}_{tk}^o}^{(-1)^{t}}}$ which is determined by the joint PMFs of sensor decisions, see Appendix\,\ref{appendix:A} for details. Also, see \eqref{eq:3logT} as an example for $L = 3$.

\subsubsection{The optimal test statistic for the three-sensor case}
Considering the three-sensor case, the joint PMF of $u_{1n}, u_{2n}$ and $u_{3n}$ at any time instant, $1 \leq n \leq N$, under $H_1$ and $H_0$ is given as follows, respectively,
\begin{equation}
\begin{aligned}
\label{eq:p}
&P(u_{1n}, u_{2n}, u_{3n}|H_1) = \\ 
&P_{000}^{(1-u_{1n})(1-u_{2n})(1-u_{3n})}\!P_{001}^{(1-u_{1n})(1-u_{2n})u_{3n}}\!P_{010}^{(1-u_{1n})u_{2n}(1-u_{3n})}  \\ 
&P_{011}^{(1-u_{1n})u_{2n}u_{3n}}\!P_{100}^{u_{1n}(1-u_{2n})(1-u_{3n})}\!P_{101}^{u_{1n}(1-u_{2n})u_{3n}} \\ 
&P_{110}^{u_{1n}u_{2n}(1-u_{3n})}\!P_{111}^{u_{1n}u_{2n}u_{3n}}, 
\end{aligned}
\end{equation}
and

\begin{equation}
\label{eq:q}
\begin{aligned}
&P(u_{1n}, u_{2n}, u_{3n}|H_0) = \\ 
&Q_{000}^{(1-u_{1n})(1-u_{2n})(1-u_{3n})}\!Q_{001}^{(1-u_{1n})(1-u_{2n})u_{3n}}\!Q_{010}^{(1-u_{1n})u_{2n}(1-u_{3n})} \\ 
&Q_{011}^{(1-u_{1n})u_{2n}u_{3n}}\!Q_{100}^{u_{1n}(1-u_{2n})(1-u_{3n})}\!Q_{101}^{u_{1n}(1-u_{2n})u_{3n}} \\ 
&Q_{110}^{u_{1n}u_{2n}(1-u_{3n})}\!Q_{111}^{u_{1n}u_{2n}u_{3n}}. 
\end{aligned}
\end{equation}

For simplification of notation, we use $A_1$ to $A_7$ to denote the coefficients of $\mathbf{u}^t$, $t = 1, 2, 3$. Substituting \eqref{eq:p} and \eqref{eq:q} into \eqref{eq:t} and taking log on both sides, we can get
\begin{equation}
\begin{aligned}
\label{eq:3logT}
&\text{log}\Lambda_1(\mathbf{u}) = \\ 
&A_1\sum_{n=1}^{N}u_{1n} + A_2\sum_{n=1}^{N}u_{2n} + A_3\sum_{n=1}^{N}u_{3n} + A_4\sum_{n=1}^{N}u_{1n}u_{2n} + \\ 
&A_5\sum_{n=1}^{N}u_{1n}u_{3n} + A_6\sum_{n=1}^{N}u_{2n}u_{3n} + A_7\sum_{n=1}^{N}u_{1n}u_{2n}u_{3n}, 
\end{aligned}
\end{equation}
where 
\begin{equation*}
\begin{aligned}
&A_1 = \text{log}\frac{Q_{000}P_{100}}{P_{000}Q_{100}}, ~~~~~~~~~~\,\, A_2 = \text{log}\frac{Q_{000}P_{010}}{P_{000}Q_{010}}, \\ 
&A_3 = \text{log}\frac{Q_{000}P_{001}}{P_{000}Q_{001}}, ~~~~~~~~~~\,\, A_4 = \text{log}\frac{P_{000}Q_{100}Q_{010} P_{110}}{Q_{000}P_{100}P_{010}Q_{110}}, \\ 
&A_5 = \text{log}\frac{P_{000}Q_{100}Q_{001}P_{101}}{Q_{000}P_{100}P_{001}Q_{101}}, A_6 = \text{log}\frac{P_{000}Q_{010}Q_{001}P_{011}}{Q_{000}P_{010}P_{001}Q_{011}}, \\
&A_7 = \text{log}\frac{Q_{000}P_{100}P_{010}P_{001}Q_{110}Q_{101}Q_{011}P_{111}}{P_{000}Q_{100}Q_{010}Q_{001}P_{110}P_{101}P_{011}Q_{111}}.
\end{aligned}
\end{equation*}

When sensor decisions among $L$ sensors are conditionally independent, only the term $\underset{\{ i_1n \} \in \mathnormal{I}_1}{\sum}A_{\mathbf{u}^1}\sum_{n = 1}^{N}\mathbf{u}^1$ in \eqref{eq:logT} is left and the optimal fusion rule reduces to the Chair-Varshney fusion rule statistic (i.e., weighted sum of sensor decisions\cite{chair1986optimal}). For correlated sensor decisions, the optimal fusion rule depends on both the weighted sum of sensor decisions and the weighted sum of the cross products of sensor decisions. The cross products of the sensor decisions are due to dependence among multiple sensors. The joint PMFs of sensor decisions, namely $P_s$ and $Q_s$, $s \in S$, determine the weights of the optimal test statistic, and can be obtained by solving $L$ integrals on the joint PDFs of the corresponding sensor observations (see the example in \eqref{eq:inte}). In the following subsection, we will propose an R-Vine copula based approach to model existing complex dependence and construct the joint PDFs of sensor observations. After obtaining the joint PMFs and given sensor decisions, the optimal fusion rule is given by
\begin{equation}
\text{log}\Lambda(\mathbf{u}) \Gtrless^{H_1}_{H_0} \gamma,
\end{equation}
where $\gamma$ is the threshold for the test at the FC.

To characterize the fusion performance at the FC using the system probabilities of detection and false alarm, we consider the asymptotic distribution of the optimal fusion rule statistic under $H_0$ and $H_1$.

\begin{theorem}
\label{theorem:testS}
The optimal fusion test statistic $\text{log}\Lambda(\mathbf{u})$ is asymptotically (when N is large) Gaussian.
\end{theorem}

The proof of the theorem and the first and second order statistics of $\text{log}\Lambda(\mathbf{u})$ under both hypotheses are given in Appendix\,\ref{appendix:testS}.

Let the first and second order statistics of $\text{log} \Lambda(\mathbf{u})$ be denoted by $\mu_0$ and $\sigma_0^2$ under $H_0$ and $\mu_1$ and $\sigma_1^2$ under $H_1$. These can be easily derived using the joint PMFs of sensor decisions. The system
probability of detection ($P_D$) and system probability of false alarm ($P_F$) are then given by
\begin{align}
P_D &= \mathnormal{Q}\left ( \frac{\gamma - \mu_1}{\sigma_1} \right),\\
P_F &= \mathnormal{Q}\left ( \frac{\gamma - \mu_0}{\sigma_0} \right),
\end{align}
where $\mathnormal{Q}(\cdot)$ is the complementary CDF of the Gaussian distribution. Under the Neyman-Pearson framework and by constraining $P_F = \alpha$, $\gamma$ can be obtained by 
\begin{equation}
\gamma = \sigma_0 \mathnormal{Q}^{-1}(P_F) + \mu_0.
\end{equation}

Note that the local sensors compress their raw measurements into binary decisions (see \eqref{eq:bq}) prior to their transmission to the FC and the corresponding sensor thresholds are assumed to be $\tau_l, l = 1, 2, \ldots, L$. Let $\boldsymbol \tau$ be the vector of sensor thresholds. Constraining $P_F = \alpha$, $P_D$ can be written as
\begin{equation}
P_D(\boldsymbol \tau) = \mathnormal{Q} \left ( \frac{\sigma_0 \mathnormal{Q}^{-1}(P_F) + \mu_0(\boldsymbol \tau) - \mu_1(\boldsymbol \tau)}{\sigma_1(\boldsymbol \tau)} \right ),
\end{equation}
where $\boldsymbol \tau$ is chosen to maximize $P_D$ at a particular value of $P_F$.

It should be noted that the computational complexity for obtaining the joint PMFs is very high since we need to perform multi-dimensional integration at each time instant. In what follows, we first propose the R-Vine copula based methodology to characterize the joint PDFs of sensor observations and then develop an efficient optimal fusion algorithm based on the R-Vine copula model. 

\subsection{R-Vine Copula Based Dependence Modeling}
According to Sklar's theorem (Section \ref{subsec:st}), the joint PDF of sensor observations can be separated into its marginals and the dependence structure that is fully characterized by the copula density (see \eqref{CopEq2}). 
As indicated earlier, the R-Vine copula model (Section \ref{subsec:vine}) is more flexible to decompose the joint PDF into its marginals and a cascade of bivariate copula densities. In the following, we will use the R-Vine copula to model the dependence structure and obtain the joint PDF of sensor observations.

In our parallel distributed detection sensor network, $L$ sensors make a set of observations $\mathbf{z}_{n} = [z_{1n}, \ldots, z_{Ln}]$ at time instant $n$. Recall that we assume the sensor observations to be conditionally i.i.d. over time. Therefore, it is sufficient to consider the joint PDF of $\mathbf{z}_{n}$. For notational convenience, we omit the index $n$ in this subsection and let $\mathbf{z} = [z_{1}, \ldots, z_{L}]$ be the $L$-dimensional observation vector with its marginal CDFs, $\mathbf F = [F_1(z_1), \ldots, F_L(z_L)]$. 
The R-Vine copula ($\mathbf F, \mathcal{V}, \mathbf B$) (see Definition \ref{def:R-VineC}) of $\mathbf{z}$ is specified by its marginal CDFs $\mathbf F$, R-Vine $\mathcal{V} = (T_1,\dots,T_{L-1})$ and a set of bivariate copulas $\mathbf B = 
\{ C_{\mathfrak{C}_{e, a}, \mathfrak{C}_{e, b} | D_e} \, | \, e \in E_i, i = 1, 2, \ldots, L - 1 \}$ with a set of parameters $\bs{\phi}$.

From \eqref{eq:densityvine}, the joint PDF of $\mathbf{z}$ is given as
\begin{align}
\label{eq:densityvine-S}
&f(\mathbf z | \mathcal{V}, \mathbf B, \bs{\phi}) = \prod_{l=1}^L f(z_{l}) \prod_{i=1}^{L-1}\prod_{e\in E_i} \times  \\ \nonumber
&c_{\mathfrak{C}_{e, a}, \mathfrak{C}_{e, b} | D_e}(F_{\mathfrak{C}_{e, a} | D_e}(z_{\mathfrak{C}_{e, a}} | \mathbf{z}_{D_e}),F_{\mathfrak{C}_{e, b} | D_e}(z_{\mathfrak{C}_{e, b}} | \mathbf{z}_{D_e}); \bs{\phi}), \nonumber 
\end{align}
where $e=\{a, b\}$, $\mathbf{z}_{D_e}=\{z_{j} | j \in D_e\}$, $f(z_{l})$ is the marginal PDF of the observation of sensor $l$, $l = 1, \dots, L$.  The conditional distributions $F_{\mathfrak{C}_{e, a} | D_e}(z_{\mathfrak{C}_{e, a}} | \mathbf{z}_{D_e})$ and $F_{\mathfrak{C}_{e, b} | D_e}(z_{\mathfrak{C}_{e, b}} | \mathbf{z}_{D_e})$ are obtained using \eqref{eq:FF}.

Given a set of $N$ observed data $\mathbf z_1, \ldots, \mathbf z_N$, the joint PDF of the observations is given as
\begin{equation}
f(\mathbf z_{1},\ldots,\mathbf z_{N}) = \prod_{n=1}^N f(\mathbf z_{n} | \mathcal{V}, \mathbf B, \bs{\phi}). 
\end{equation}

\subsection{Model Selection and Estimation}
\label{subsec:Model}
The fitting of an R-Vine copula model to given data requires the selection of the R-Vine tree structure $\mathcal{V}$, the choice of copula families for the bivariate copula set $\mathbf B$ and the estimation of their corresponding parameters $\bs{\phi}$. Since the bivariate copula families and their corresponding parameters both depend on the R-Vine tree structure, the identification of trees accurately is key to the R-Vine copula model. It has been shown that the number of possible R-Vines for $n$ variables increases very rapidly and is given by $\binom{n}{2} \times (n - 2)! \times 2^{\binom{n-2}{2}}$ \cite{morales2010number}. It is not computationally feasible to find the best model by fitting all possible R-Vine constructions. Suboptimal R-Vine copula selection strategies have been investigated in the literature. In \cite{dissmann2013selecting}, a sequential method to select an R-Vine model based on Kendall's tau was proposed, where a maximum spanning tree algorithm was used. Moreover, the feasibility and efficiency of this method was demonstrated. The sequential method starts with the selection of the first tree $T_1$ and continues tree by tree up to the last tree $T_{L-1}$. The trees are selected in a way that the chosen bivariate copula models the strongest pair-wise dependencies present which are characterized by Kendall's tau. 
There are other possible choices to measure the pair-wise dependencies besides Kendall's tau, for example, the Akaike Information Criterion (AIC) \cite{akaike1973information} of each bivariate copula proposed in \cite{czado2013selection} and the $p$-value of a copula goodness of fit test and variants proposed in \cite{czado2013selection2}. 

In this paper, we adopt the sequential method proposed in \cite{dissmann2013selecting} to construct the R-Vine copula model. Also, we use Kendall's tau as the measure of dependencies and select the spanning tree that maximizes the sum of the absolute values of empirical Kendall's tau. Kendall's tau can be expressed as an expectation over a bivariate copula distribution as shown in \cite{nelsen2013introduction}, and typically, the log likelihood of a bivariate copula increases with increasing absolute values of Kendall's tau. Moreover, the advantage of using Kendall's tau is that one does not need to select and estimate the bivariate copulas prior to the tree selection step. We summarize the sequential method based on Kendall's tau for obtaining the joint PDF of sensor observations in Algorithm \ref{algo1}, where the weights $w_{i,j}$ denote the absolute values of the empirical Kendall's tau and the trees are selected sequentially by maximizing the sum of the absolute values of empirical Kendall's taus. After the tree structure is determined, we select the best copulas for each pair of variables from the defined copula library. At the end, we obtain the R-Vine density function. The selection of the best copulas and the estimation of their corresponding parameters are presented in the following.

Besides the selection of the R-Vine tree structure, we need to define a copula family for each pair of sensors and select the copula that best characterizes the pair-wise dependencies. Consider a library of copulas, $\mathcal{C} =  \{c_m: m=1,\ldots,M\}$ and assume that we have a set of $N$ observations $\mathbf z_1, \ldots, \mathbf z_N$. Based on \eqref{eq:densityvine-S}, to obtain the joint PDF of sensor observations, we need to specify the marginal PDFs, marginal CDFs including conditional marginal CDFs of individual local sensor observations as well as the bivariate dependence structure. If we do not have any prior knowledge of the phenomenon of interest, the marginal PDFs $f(z_{ln})$ for sensor $l, l = 1, 2, \ldots, L$ at time instant $n, n = 1, 2, \ldots, N$ can be estimated non-parametrically using Kernel density estimators \cite{wassermann2006all}, and the marginal CDFs $F(z_{ln})$ can be determined by the Empirical Probability Integral Transforms (EPIT) \cite{he2012fusing}. Note that the conditional marginal CDFs need to be obtained recursively using \eqref{eq:FF}. Before selecting the best bivariate copula, the copula parameter set $\bs{\phi}$ is obtained using MLE, which is given by 
\begin{equation}
\label{eq:phi}
\widehat{\bs{\phi}} = \text{arg}\max_{ \bs{\phi} } \sum_{n=1}^N{ \log c(F(z_{l_1n}), F(z_{l_2n})|\bs{\phi}) },
\end{equation}
where $(l_1, l_2), l_1, l_2 \in [1, 2, \ldots, L]$ is a connected pair in R-Vine tree $\mathcal{V}$ and for simplification of notation, we omit the conditioned elements for conditional marginal CDFs. 

To decide on the best copula, we consider three widely used model selection criteria: AIC, Bayesian Information Criterion (BIC) \cite{schwarz1978estimating}, and MLE,
\begin{equation}
\begin{aligned}
&\text{AIC} = - \sum_{n = 1}^{N} \log c(F(z_{l_1n}), F(z_{l_2n})|\widehat{\bs{\phi}}) + 2q(L), \\
&\text{BIC} = - \sum_{n = 1}^{N} \log c(F(z_{l_1n}), F(z_{l_2n})|\widehat{\bs{\phi}}) + q(L) \log(N), \\
&\text{MLE} =  \sum_{n = 1}^{N} \log c(F(z_{l_1n}), F(z_{l_2n})|\widehat{\bs{\phi}}), 
\end{aligned}
\end{equation}
where $q(L)$ is the number of parameters in the R-Vine model and $N$ is the number of observations.

\subsection{Efficient R-Vine Copula Based Fusion with Correlated Decisions}
As observed in the optimal test statistic \eqref{eq:PQ}, the set of joint PMFs $P_s$ and $Q_s, s \in S$ are required to be obtained at each time instant. To tackle the computational complexity resulting from multi-dimensional integration, we propose an efficient approach for R-Vine copula based fusion of correlated decisions.

Let the local sensor probability of detection and local sensor probability of false alarm be represented by $p_{l}$ and $q_{l}$ for sensor $l, l = 1, 2, \ldots, L$. Therefore, $p_{l}$ and $q_{l}$ are given as
\begin{equation}
\label{eq:pq}
\begin{aligned}
p_{l} &=  {\int_{\tau_{l}}^{+\infty}}\mathnormal{f}(z_{l}|H_1)\mathnormal{d}{z_{l}},\\
q_{l} &= {\int_{\tau_{l}}^{+\infty}}\mathnormal{f}(z_{l}|H_0)\mathnormal{d}{z_{l}},
\end{aligned}
\end{equation}
where $\tau_{l}$ is the quantization threshold for sensor $l$. The local optimal sensor thresholds under the Neyman-Pearson criterion are obtained by solving the following problem:
\begin{align}
\label{eq:thre}
\begin{array}{ll}
\displaystyle \underset{\tau_{l}}{\text{maximize}} &  p_{l},   \\
 \text{subject to} & q_{l} \leq \beta_l,
\end{array}
\end{align}
where $\beta_l$ is the constraint on the local probability of false alarm for sensor $l$, $p_l$ and $q_l$ are given in \eqref{eq:pq}.

Consider the set of joint PMFs under hypothesis $H_1$, namely $P_s, s \in S$.
Let $\tilde{A}_l = \{u_1u_2 \ldots u_l \ldots u_L | u_l = 0\}$ and $\tilde{A}_l^c$ denote the complement of $\tilde{A}_l$  for $ l = 1, 2, \ldots, L$. Note that the union of the sets $\tilde{A}_1, \tilde{A}_2, \ldots, \tilde{A}_L$ is $S$.
For the three-sensor case, we have $\tilde{A}_1  = \{\{011\}, \{010\}, \{001\}, \{000\} \}$,  $\tilde{A}_2  = \{\{101\}, \{100\}, \{001\}, \{000\} \}$ and $\tilde{A}_3  = \{\{110\}, \{100\}, \{010\}, \{000\} \}$. For any $s \in S$, the PMF under hypothesis $H_1$  is given as
\begin{align}
\label{eq:P}
P_s = P(\bigcap_{l = 1} ^{L}B_l), 
\end{align}
where $B_l = \tilde{A}_l$ if $s_l = 0$, otherwise, $B_l = \tilde{A}_l^c$. $P_s$ can be obtained using copula functions. For example, $P_{101}$ is given as
\begin{align}
\label{eq:P2}
P_{101} &= P(\tilde{A}_1^c \cap \tilde{A}_2 \cap \tilde{A}_3^c) \\ \nonumber
&= P(\tilde{A}_2 - \tilde{A}_2 \cap \tilde{A}_3 - \tilde{A}_1 \cap \tilde{A}_2 + \tilde{A}_1 \cap \tilde{A}_2 \cap \tilde{A}_3) \\ \nonumber
&= 1 - p_2 - C_{23}(1 - p_2, 1 - p_3) - C_{12}(1 - p_1, 1 - p_2) \\ \nonumber
&+ C_{123}(1 - p_1, 1 - p_2, 1 - p_3), \nonumber
\end{align}
where $C_{12}$, $C_{23}$ and $C_{123}$ are copula functions.

Consider the three-sensor case, the joint PMFs under $H_1$ is given as
\begin{align*}
&P(u_1 = 0, u_2 = 0, u_3 = 0) = C_{123}\\ \nonumber
&P(u_1 = 0, u_2 = 0, u_3 = 1) = C_{12} - C_{123}\\ \nonumber
&P(u_1 = 0, u_2 = 1, u_3 = 0) = C_{13} - C_{123}\\ \nonumber
&P(u_1 = 0, u_2 = 1, u_3 = 1) = 1 - p_1 - C_{12} - C_{13} + C_{123}\\ \nonumber
&P(u_1 = 1, u_2 = 0, u_3 = 0) = C_{23} - C_{123}\\ \nonumber
&P(u_1 = 1, u_2 = 0, u_3 = 1) = 1 - p_2 - C_{12} - C_{23} + C_{123}\\ \nonumber
&P(u_1 = 1, u_2 = 1, u_3 = 0) = 1 - p_3 - C_{23} - C_{13} + C_{123} \nonumber
\end{align*}
where we omit the marginal CDFs of $C$, namely $1 - p_l, l = 1, 2, \ldots, L$. Similarly, PMFs under $H_0$ are obtained with $p_l$ replaced by $q_l$, $l = 1, 2, \ldots, L$.

Define $\mathcal{C}$ as the set that specifies all the copula functions involved in the PMFs of sensor decisions. We further define the index set of $\mathcal{C}$ as $G$ which is the union of all the nonempty subsets with at least two elements of set $\{ 1, 2, \ldots, L \}$ in sorted order and the cardinality of set $G$ is $\left | G \right | = N_G = \sum_{k=2}^{L}\binom{L}{k}$. For the three-sensor case, we have the copula function set $\mathcal{C} = \{ C_{12}, C_{13}, C_{23}, C_{123} \}$ and its index set $G = \{ \{1, 2\}, \{1, 3\}, \{2, 3\}, \{1, 2, 3\} \}$. 

As we can see, knowing $\mathcal{C}$, we can obtain all combinations of the joint PMFs.  Any arbitrary copula density function of $C \in \mathcal{C}$ can be obtained through Algorithm \ref{algo1}. By integrating the copula density function, we can obtain the copula function $C \in \mathcal{C}$. The computation is significantly reduced using the copula function set $\mathcal{C}$ to obtain the joint PMFs since we only need to perform multi-dimensional integration once for each copula function $C \in \mathcal{C}$. To further reduce computational complexity, we start with $L$-dimensional R-Vine copula model selection by applying Algorithm \ref{algo1} and then use the obtained optimal tree structure with its R-Vine matrix $\mathbf M^*$ (see Definition \ref{def:R-VArr}), R-Vine copula family matrix $\mathbf F^*$ and the corresponding parameter matrix $P^*$ to directly get the copula density functions that need to be estimated in $\mathcal{C}$. For the rest of the copula functions to be estimated, we again start with selecting an appropriate R-Vine copula model with largest dimension and use its optimal tree structure to obtain lower dimensional copula functions that have not been estimated. We proceed with this procedure till we obtain all the copula functions in the set $\mathcal{C}$.  For the R-Vine copula example  in Fig.\,\ref{fig: example_RVine}, from its R-Vine matrix $\mathbf M^*$ (see Section\, \ref{sec: CT}) with its optimal R-Vine copula family matrix and the corresponding parameter matrix, we can directly obtain the density of $c_{35}, c_{24}, c_{12}, c_{23}, c_{123}, c_{1234}, c_{12345}$.

The proposed efficient optimal fusion rule is summarized in Algorithm \ref{algo2}.


\begin{algorithm}
	Inputs: Marginal PDFs of local sensor observations $f(z_{i} | H_1)$ for sensor $i$, $i = 1, 2, \ldots, m, m \in [1, 2, \ldots, L]$, a predefined copula library $\mathcal{C}$.
	\begin{enumerate}
	         \item Get marginal CDFs of local sensor observations $F_i$, $i = 1, 2, \ldots, m$.
		\item Calculate the weight $\mathbf{w}_{i,j}$ for all possible pairs of sensors $\{ i, j \}$, $1 \leq i \leq j \leq m$.
		\item Select the maximum spanning tree, i.e.,
		\begin{equation*}
		T_1 = \underset{T = (M, E) \; \text{spanning tree}}{\text{argmax}} \underset{e \in E}{\mathbf{\sum}}\mathbf{w}_{i(e), j(e)}.
		\end{equation*}
		\item For each edge $e \in E_1$, select a copula $C_{i(e), j(e)}$ and estimate the corresponding parameter(s) $\bs{\phi}_{i(e), j(e)}$.
		\item Obtain $F_{i(e) | j(e)}(z_{i(e)} | z_{j(e)})$ and $F_{j(e) | i(e)}(z_{j(e)} | z_{i(e)})$ using \eqref{eq:FF}.
		\item For $s = 2, \ldots, m-1$, calculate the weight $\mathbf{w}_{i(e), j(e) | D(e)}$ for all conditional variable pairs $\{ {i(e), j(e) | D(e)} \}$.
		\item Among these edges, select the maximum spanning tree, i.e.,
		\begin{equation*}
		T_s = \underset{T = (M, E) \; \text{spanning tree with}\; E \subset E_p}{\text{argmax}} \underset{e \in E}{\mathbf{\sum}}\mathbf{w}_{i(e), j(e) | D(e)}.
		\end{equation*}
		\item For each edge $e \in E_s$, select a conditional copula $C_{i(e), j(e) | D(e)}$ and estimate the corresponding parameters $\bs{\phi}_{i(e), j(e) | D(e)}$.
		\item Obtain $F_{i(e) | j(e) \cup D(e)}(z_{i(e)n} | z_{j(e)n}, z_{D(e)})$ and $F_{i(e) | j(e) \cup D(e)}(z_{j(e)} | z_{i(e)}, z_{D(e)})$ using \eqref{eq:FF}.
		\item Obtain the R-Vine copula density $c$.
		\item Obtain the joint PDF of sensor observations using \eqref{eq:densityvine-S}.
		\end{enumerate}
	\caption{Sequential method to obtain the joint PDF of sensor observations.}
	\label{algo1}
\end{algorithm} 

\begin{algorithm}
Inputs: Marginal PDFs of local sensor observations $f(z_{l} | H_1)$, $l = 1, 2, \ldots, L$.
\begin{enumerate}
	        \item Obtain optimal local quantizer threshold $\mathbf{\tau}_l, l = 1, 2, \ldots, L$ for all sensors by solving problem \eqref{eq:thre}.
	        \item Calculate local sensor probability of detection $p_l$ and probability of false alarm $q_l$ for all sensors, $l = 1, 2, \ldots, L$.
	        \item Obtain optimal R-Vine structure of $L$ sensors using algorithm \ref{algo1} and its R-Vine matrix $\mathbf{M}^*$ and the corresponding R-Vine copula family matrix $\mathbf{F}^*$ and parameter matrix $\mathbf{P}^*$.
	        \item For $i = 1, 2, \ldots, L-1$,
	        \begin{enumerate}
	        \item Let $G_1 = \emptyset$.
	        \item Obtain $C_{\mathbf{M}^*_{i, i}, \mathbf{M}^*_{L, i}}$ and $C_{\mathbf{M}^*_{i, i}, \mathbf{M}^*_{i+1, i}, \ldots, \mathcal{M}^*_{L, i}}$ directly from the obtained R-Vine copula family matrix $\mathbf{F}^*$ and parameter matrix $\mathbf{P}^*$.
	        \item $G_1 =  G_1 \cup \{ \{ \mathbf{M}^*_{i, i}, \mathbf{M}^*_{L, i} \}, \{\mathbf{M}^*_{i, i}, \mathbf{M}^*_{i+1, i}, \ldots, \mathbf{M}^*_{L, i}\} \}$.
	        \end{enumerate}
	        \item For $g = 1, 2, \ldots, N_G-1$,
	        \begin{enumerate}
	        \item if $G(g) \neq a, \forall a \in G_1$.
	        \item Apply algorithm \ref{algo1} and obtain $C_{G(g)}$.
	        \end{enumerate}
		\item Calculate the PMFs of sensor decisions under hypotheses $H_1$ and $H_0$, respectively, using \eqref{eq:P}.
		\item Solve the detection testing problem \eqref{eq:logT}.
\end{enumerate}
\caption{Efficient optimal fusion rule.}
\label{algo2}
\end{algorithm}

\section{Numerical Results}
\label{sec:nr}
In this section, we demonstrate the efficacy of our proposed R-Vine copula based fusion methodology for the problem of distributed detection through numerical examples. We assume that there are two hypotheses, where $H_1$ denotes the presence of a signal $s$ and $H_0$ indicates the absence of $s$. In the distributed sensor network we consider in this paper, we assume that three sensors sense and acquire raw measurements of the signal $s$ via a linear sensing model, and then quantize the detected signal into a single-bit local decision. After compression, the decisions are transmitted to the FC. The signals received at the sensors can be modeled as:
\begin{align}
&H_1: z_{in} = h_{in} s_{in} + w_{in}, &i = 1, 2, 3; n = 1, \ldots, N \\ \nonumber
&H_0: z_{in} = w_{in}, &i = 1, 2, 3; n = 1, \ldots, N \nonumber
\end{align}
where $z_{in}$, $h_{in}$ and $w_{in}$ denote the received signal, the fading channel gain and the measurement noise at sensor $i$ and time instant $n$. Moreover, $s_{in}$ is the target signal received by the $i$th sensor at $n$th time instant. The intensity of the signal $s$ is assumed to be a constant. We assume that the channel gain $h_{in}$ is chosen randomly and independently from $\mathrm{Rayleigh}(\xi)$ distribution with parameter $\xi$ over time. 
However, $h_{in}$ can be spatially dependent. The measurement noise $w_{in}$ is drawn from zero-mean Gaussian distribution with standard deviation $\sigma_w$ ($\sigma_{w1} = 1$, $\sigma_{w2} = 0.9$  and $\sigma_{w3} = 0.8$) and is assumed to be temporally independent conditioned on either hypothesis but can be spatially correlated. Furthermore, we assume that the measurement noise, the fading gains, and the target signal are mutually independent. Also, we assume that we do not have any prior knowledge of the marginals and dependence structure. Unless specified otherwise, the number of sensor observations is assumed to be $N = 100$, the local probability of false alarm is constrained by $q_l \leq 0.1, l = 1, 2, 3$ and AIC is used for optimal bivariate copula selection. 

To demonstrate the superiority of R-Vine copula, we apply standard multivariate copula and seven different R-Vine classes given by
\begin{enumerate}
\item Mixed R-Vine: R-Vine with pair-copula terms chosen individually from 15 bivariate copula types (Gauss, Student-t, Gumbel, Clayton, Frank and Joe etc.).
\item all Gaussian R-Vine: R-Vine with each pair-copula term chosen as bivariate Gaussian copula.
\item all Student $t$ R-Vine: R-Vine with each pair-copula term chosen as bivariate Student $t$ copula.
\item all Gumbel R-Vine: R-Vine with each pair-copula term chosen as bivariate Gumbel copula.
\item all Clayton R-Vine: R-Vine with each pair-copula term chosen as bivariate Clayton copula.
\item all Frank R-Vine: R-Vine with each pair-copula term chosen as bivariate Frank copula.
\item all Joe R-Vine: R-Vine with each pair-copula term chosen as bivariate Joe copula.
\end{enumerate}

\begin{table}[thb]
\centering
\caption{The performance of R-Vine classes and standard multivariate copulas.}
\label{table:eq}
\begin{tabular}{|c|c|c|c|c|}
\hline
 & MLE  & AIC & BIC & $p$-value                 \\ \hline
R-Vine mixed           & 6300.72 & -12595.44  &-12575.88  & 0.92              \\ \hline
R-Vine all Gaussian            & 4572.36 & -9138.72  & -9119.16 & 0.48  \\ \hline
R-Vine all Student $t$             & 4868.76 & -9725.52  &-9686.42 &0.38        \\ \hline
R-Vine all Gumbel             & 5799.94 & -11593.87  &-11574.32 & 0.57           \\ \hline
R-Vine all Clayton            & 6161.90 & -12317.8  &-12298.25 &0.82           \\ \hline
R-Vine all Frank            & 4553.14 & -9100.29  &-9080.74 &0.57           \\ \hline
R-Vine all Joe             & 6130.61 & -12255.22  &-12235.67 & 0.74           \\ \hline
Multi-Clayton copula & && & 0.0005  \\ \hline
Multi-Gaussian copula & && & 0.0005  \\ \hline
Multi-Frank copula & && & 0.0005  \\ \hline
\end{tabular}
\end{table}

Performing a parametric bootstrap with repetition rate $B = 1000$ and sample size $N = 5000$, the goodness-of-fit test results (see Table \ref{table:eq}) confirm that the R-Vine mixed model (the optimal fusion methodology) can not be rejected at a $5\%$ significance level, i.e., that the R-Vine mixed model fits the data quite well. The R-Vine models with a single type of bivariate copulas have a smaller significance than the R-Vine mixed model. The standard multivariate copulas, e.g., multivariate Clayton, Gaussian and Frank copulas, are rejected at a $5\%$ significance level. This indicates that the standard multivariate copulas are quite limited in their ability to characterize complex dependence.

To exhibit the performance improvement by applying R-Vine copula based fusion of correlated sensor decisions, we also evaluate the detection performance obtained by using 
the Chair-Varshney fusion rule that assumes independence of sensor decisions. 
Here, the R-Vine copula based fusion rule is obtained by choosing from $40$ bivariate copula types. We use receiver operating characteristics (ROCs) to characterize the detection performance. The orange diagonal curves in each ROC figure denote the performance of the random guess detector. For clarity, we summarize the empirically studied cases as follows.

\begin{itemize}
\item Case 1: We assume that the fading channel gains are spatially dependent. The measurement noises and the target signals received at the local sensors are assumed to be spatially and temporally independent. 
\item Case 2: We assume that the target signals received at the local sensors are spatially dependent but are assumed to be temporally independent conditioned on either hypothesis. The measurement noises are assumed to be spatially and temporally independent. To characterize the performance of this case, we further assume that the channels are ideal.
\item Case 3: We assume that the measurement noises are spatially dependent. The target signals received at the local sensors are assumed to be spatially and temporally independent and the channels are ideal . 
\end{itemize}


\begin{figure}[htb]
\centering
\includegraphics[width=7cm]{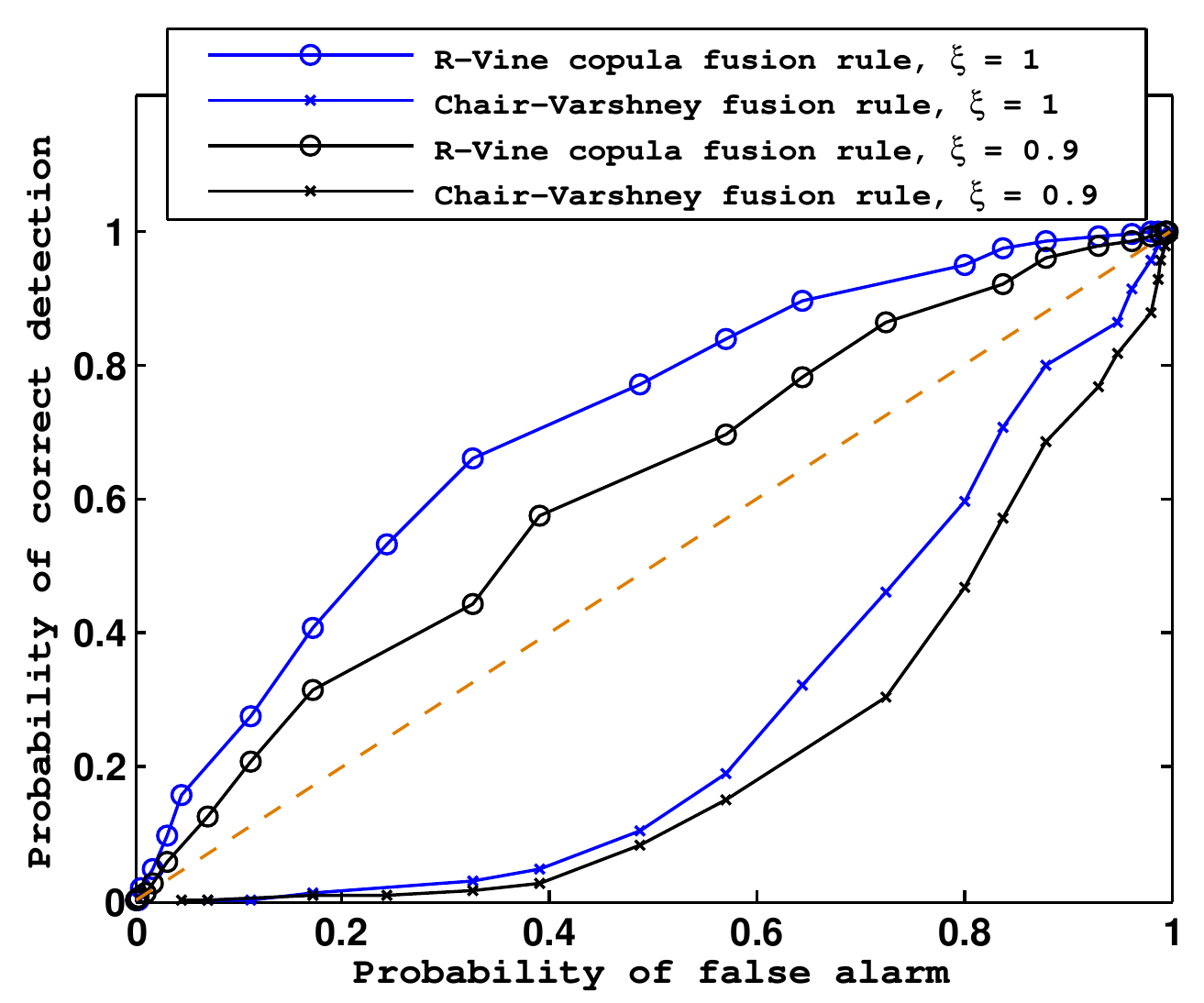} 
\caption{\footnotesize{ROCs comparing Chair-Varshney fusion rule and R-Vine copula based fusion rule with correlated fading channels.}}
\label{fig:channel}
\end{figure}

\begin{figure}[ht!]
\centering
\includegraphics[width=7cm]{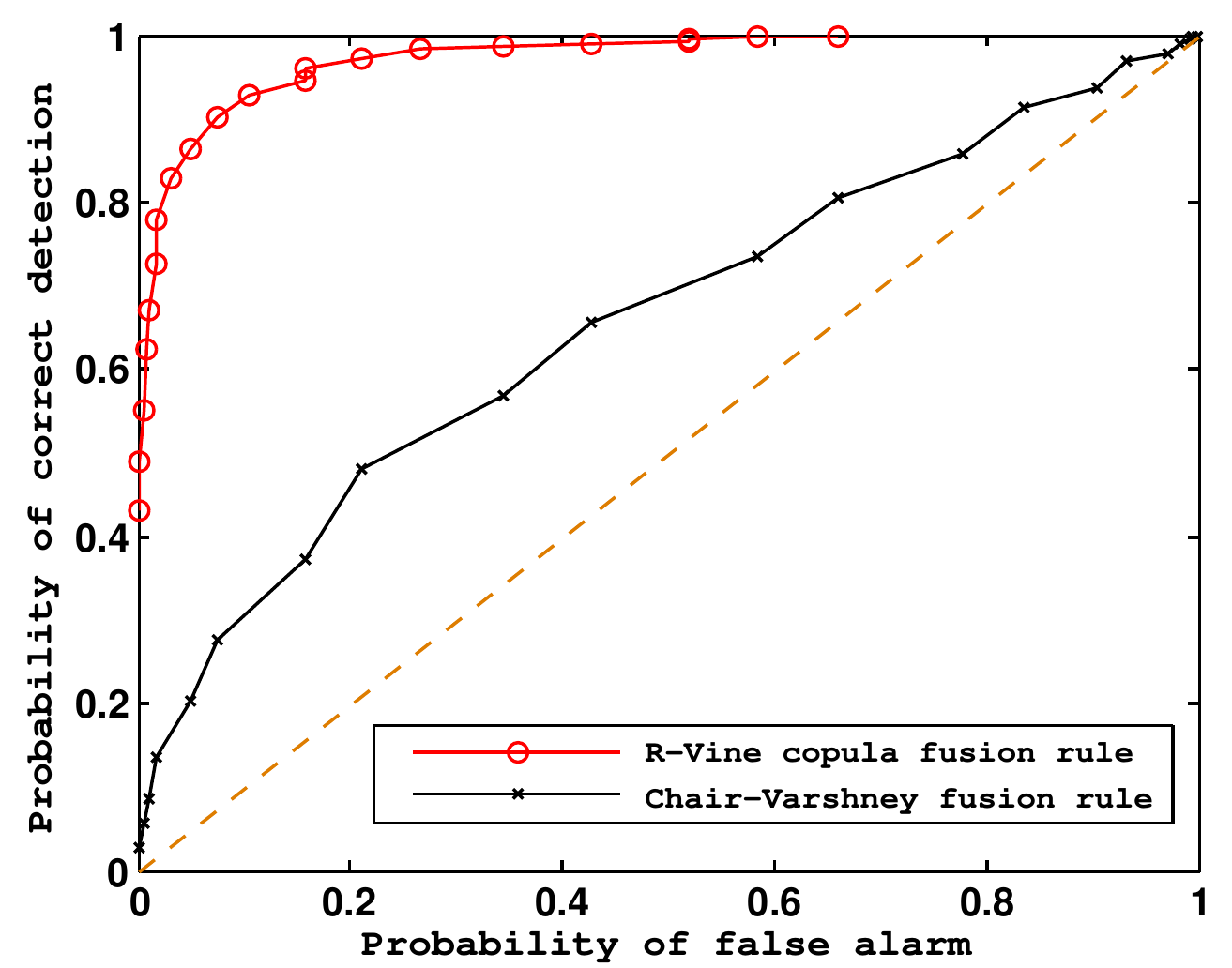} 
\caption{\footnotesize{ROCs comparing Chair-Varshney fusion rule and R-Vine copula based fusion rule with correlated signals.}}
\label{fig:signal2}
\end{figure}

\begin{figure}[ht!]
\centering
\includegraphics[width=7cm]{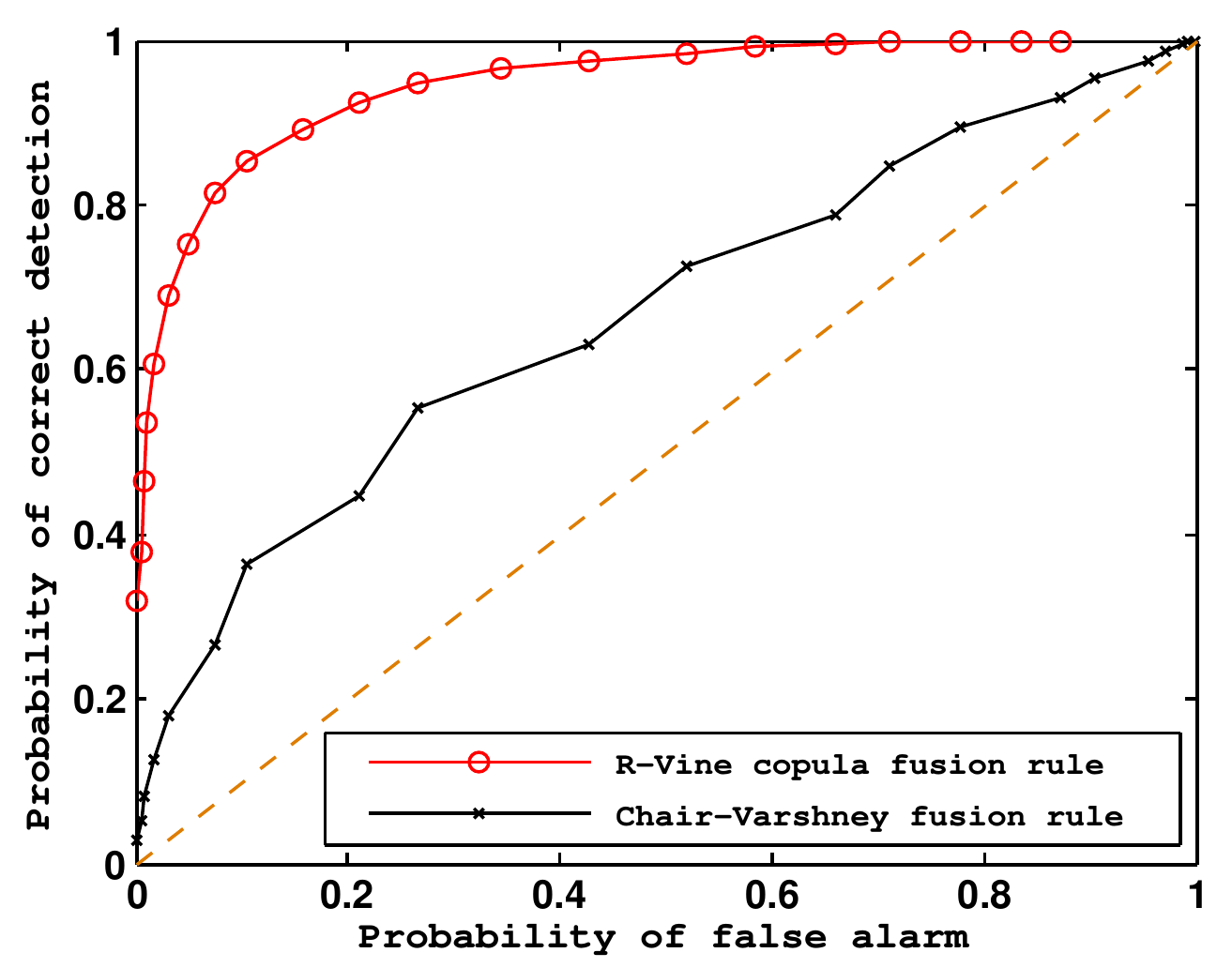} 
\caption{\footnotesize{ROCs comparing Chair-Varshney fusion rule and R-Vine copula based fusion rule with correlated signals for weaker dependence.}}
\label{fig:signal3}
\end{figure}

\begin{figure}[ht!]
\centering
\includegraphics[width=7cm]{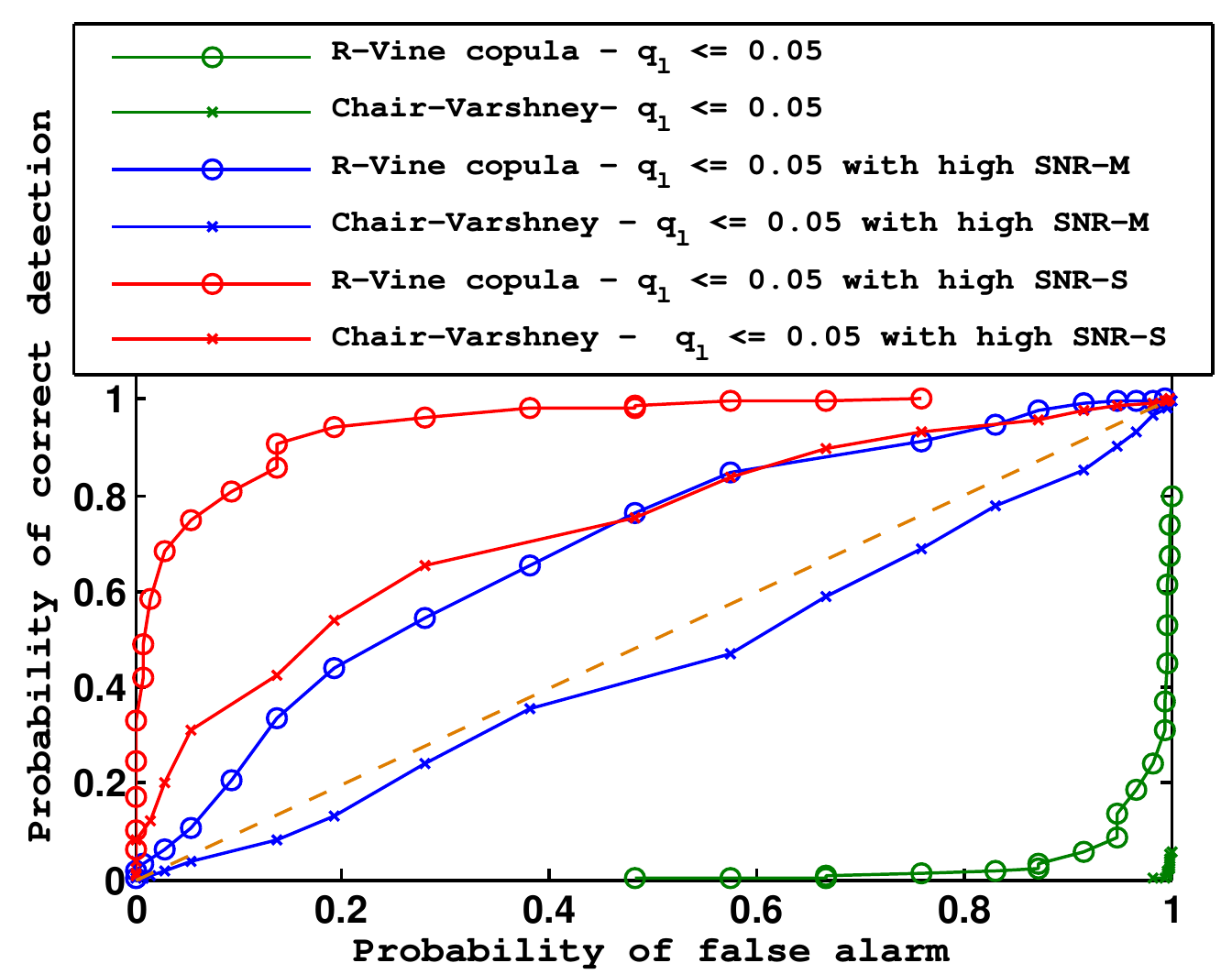} 
\caption{\footnotesize{ROCs comparing Chair-Varshney fusion rule and R-Vine copula based fusion rule with correlated signals for $q_l \leq 0.05$.}}
\label{fig:signal4}
\end{figure}

\begin{figure}[ht!]
\centering
\includegraphics[width=7cm]{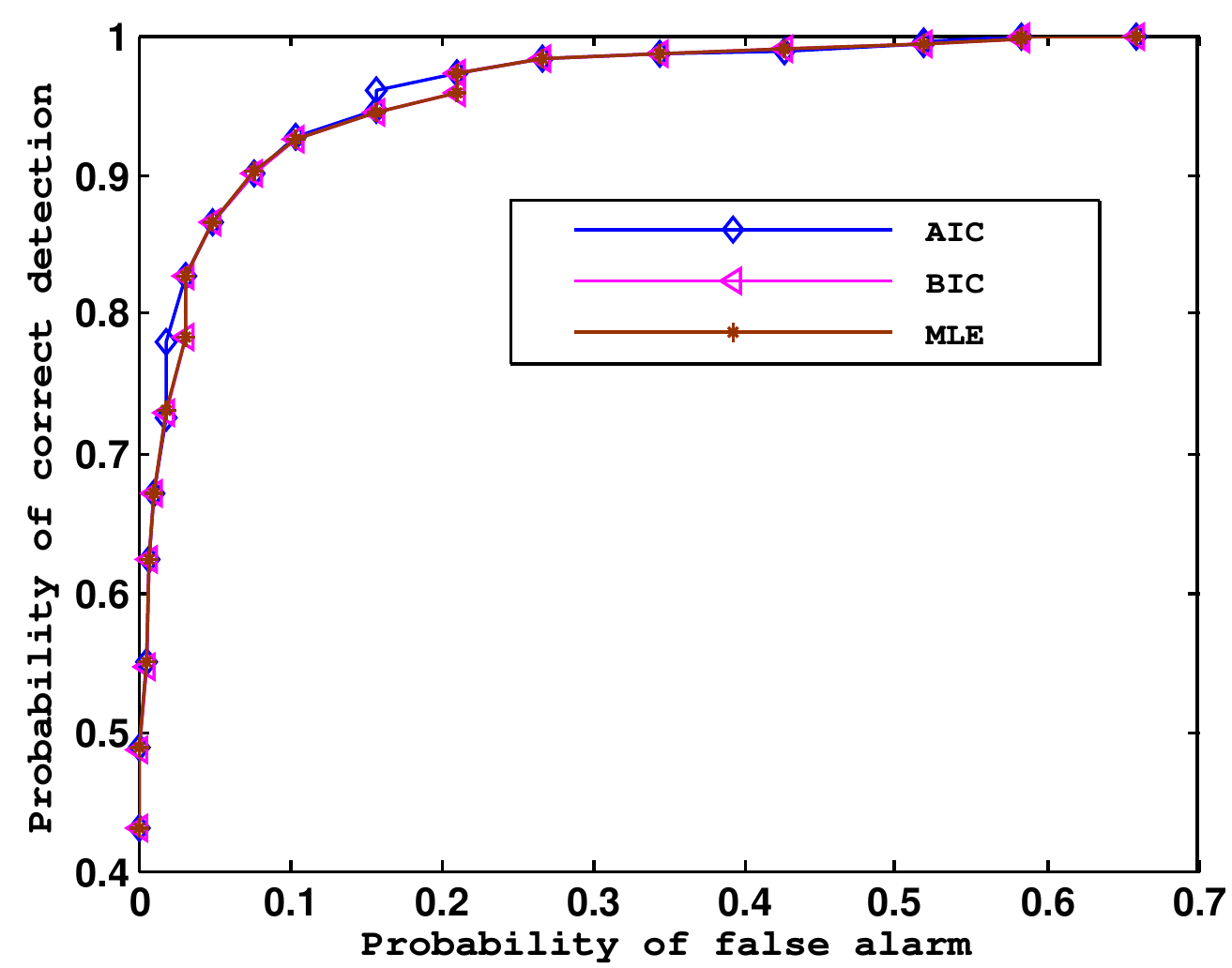} 
\caption{\footnotesize{ROCs for R-Vine copula based fusion rule with correlated signals for three model selection criteria.}}
\label{fig:signal5}
\end{figure}

\begin{figure}[ht!]
\centering
\includegraphics[width=7cm]{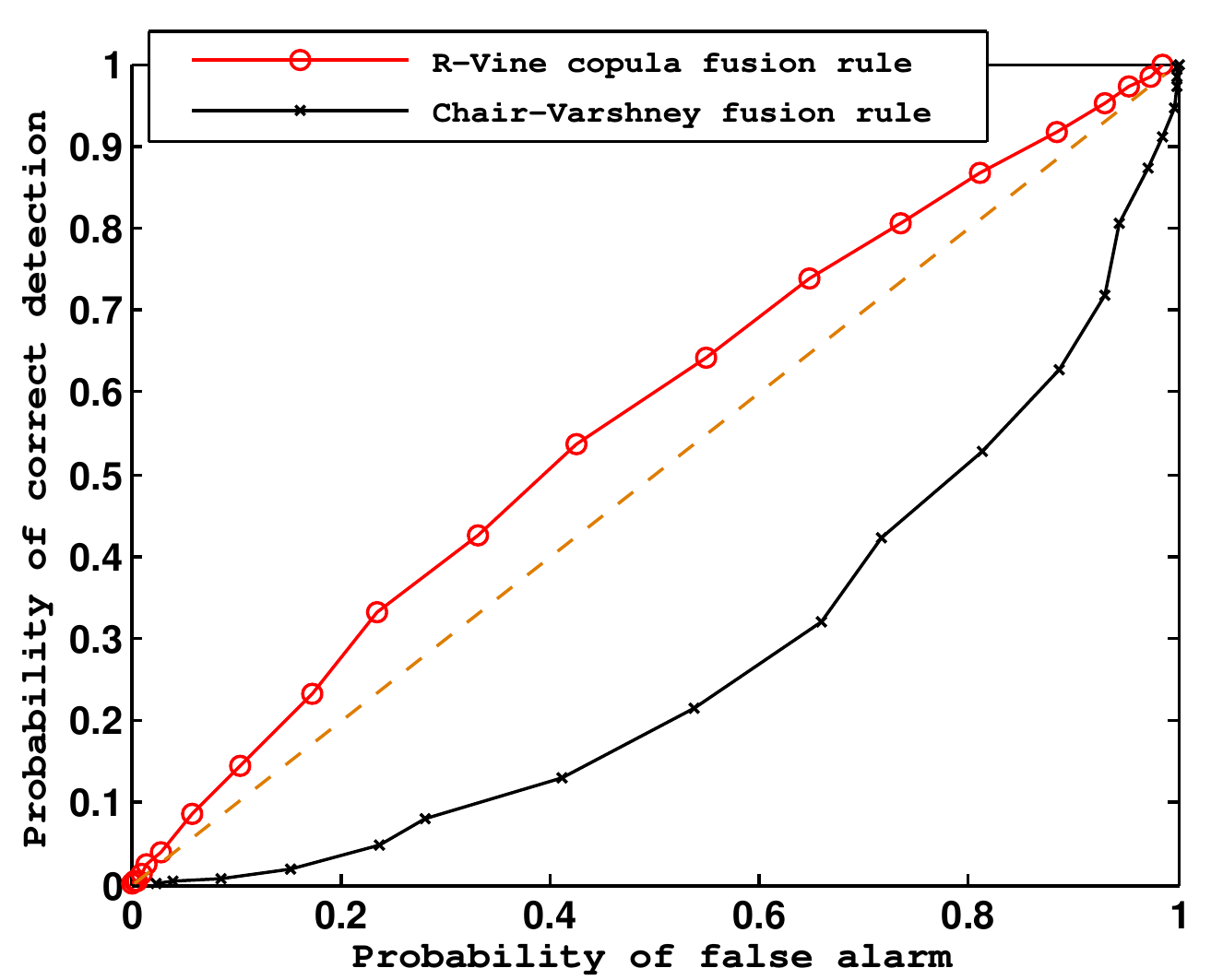} 
\caption{\footnotesize{ROCs comparing Chair-Varshney fusion rule and R-Vine copula based fusion rule with correlated measurement noise.}}
\label{fig:noise1}
\end{figure}


In Fig. \ref{fig:channel}, we present the ROCs comparing the two fusion rules: Chair-Varshney fusion rule and the proposed R-Vine copula based fusion rule for case $1$ with different fading scales, $\xi$. The intensity of the signal at the local sensors is assumed to be $s_i = 4, i = 1, 2, 3$. As we can see, the detection performance of the R-Vine copula based fusion rule is significantly better than that of the Chair-Varshney fusion rule. Moreover, with stronger fading ($\xi = 0.9$),
we can see that the detection performance is degraded compared to the fading with scale $\xi = 1$. 

In Fig. \ref{fig:signal2} and Fig. \ref{fig:signal3}, we give the ROCs comparing the Chair-Varshney fusion rule and the proposed R-Vine copula based fusion rule for Case $2$ under different dependence structures. The intensity of the signal received at the local sensors is assumed to be $s_i = 2.4, i = 1, 2, 3$. Fig. \ref{fig:signal2} shows the detection performance under a strong dependence structure and  Fig. \ref{fig:signal3} gives the detection performance under a weaker dependence structure. As we can see, for both scenarios, the detection performance of the R-Vine copula based fusion rule is significantly better than that of the Chair-Varshney fusion rule. We further show the ROCs with the local probability of false alarm constrained by $q_l \leq 0.05, l = 1, 2, 3$ in Fig. \ref{fig:signal4}. We can see that it is very difficult to detect the presence of the target signal for both the fusion rules as we have more tight false alarm constraints. By increasing the intensity of the signal to be $s_i = 3, i = 1, 2, 3$, namely with high signal to noise ratio (SNR) in terms of stronger signal power  (denoted by SNR-S) or decreasing the standard deviation of the measurement noise to be $\sigma_{wi} = 0.7, i = 1, 2, 3$, namely with high SNR in terms of weaker measurement noise power (denoted by SNR-M),  we can see that the detection performance is much better compared to weaker signal intensity or stronger measurement noise cases.


In Fig. \ref{fig:signal5}, we show the ROCs comparing the different model selection criteria discussed in Section\,\ref{subsec:Model}, namely, AIC, BIC and MLE for the proposed R-Vine copula based fusion rule. As we observe, the three criteria perform very well. The AIC criterion performs slightly better than the BIC and MLE.  

In Fig. \ref{fig:noise1}, we present the ROCs comparing the Chair-Varshney fusion rule and the proposed R-Vine copula based fusion rule for Case $3$. As expected, the detection performance of the R-Vine copula based fusion rule is much superior to that of the Chair-Varshney fusion rule. 

\begin{remark}
Note that the ROC curves that are below the diagonal indicate that the models used are not appropriate models \cite{flach2005repairing, fawcett2006introduction}. In this case, it implies that with highly correlated data, the Chair-Varshney rule (the optimum rule under independence assumption) is not able to characterize the joint statistics of our system and the random guess detector works better in this scenario. Therefore, we need better models which can give us the global information via sensor observations. In our previous paper \cite{he2012fusing}, for highly correlated observations, we had similar ROC curves.
\end{remark}

\begin{remark}
Note that in Fig. \ref{fig:noise1}, even with the R-Vine copula method, we are able to achieve performance that is only slightly better than the random guess detector. It is due to the detector operating in a difficult environment (the dependence structure in this case is quite complex) rather than the limitation of the method. Having said that, there may be other more complex copula based models that can improve system performance further.
\end{remark}

\section{Conclusion and Discussion}
\label{sec: conclusion}
In this paper, we studied the problem of distributed detection with dependent sensor decisions. We proposed a novel and powerful methodology to fuse dependent decisions obtained by binary quantization of correlated sensor observations under the Neyman-Pearson framework. To derive the optimal fusion rule, we used the R-Vine copula model to characterize the complex dependence among multiple sensors. The proposed R-Vine copula based fusion methodology was employed to overcome the limitation of the existing standard multivariate copulas, and since this methodology is extremely flexible to model complex dependence structures. The optimal log likelihood test statistics at the FC involves multi-dimensional integration at each time, leading to very high computational complexity. We proposed an efficient R-Vine copula based optimal fusion algorithm. Numerical results have illustrated the efficiency of our approach. 

In future work, one can generalize the regular vine copula model to a mixture of the regular vine copula models to find hidden dependence structures. Also, one can study multi-bit quantization at local detectors. Lastly, an efficient approach to select the global optimal regular vine tree could be taken into account.

\appendices
\section{Standard Multivariate Copula Functions}
\label{appendix:copula}
\subsection{Gaussian copula}
The multivariate Gaussian copula, derived from a multivariate Gaussian distribution, is defined as
\begin{equation}
C^{G}(\mathbf{u}|\Sigma) = \Phi_{\Sigma}(\Phi^{-1}(u_1), \ldots, \Phi^{-1}(u_L)),
\end{equation}
where $\Sigma$ is the correlation matrix, $\Phi$ is the univariate normal CDF and $\Phi_{\Sigma}$ denotes the multivariate normal CDF. 
\subsection{Student-$t$ copula}
The Student-$t$ copula is derived from a multivariate Student-$t$ distribution, which is given by
\begin{equation}
C^{t}(\mathbf{u}|\Sigma,\nu) = t_{\nu,\Sigma}(t_{\nu}^{-1}(u_1), \ldots, t_{\nu}^{-1}(u_L)),
\end{equation}
where $t_{\nu,\Sigma}$ denotes the multivariate Student-$t$ distribution with correlation matrix $\Sigma$ and degrees of freedom $\nu$ ($\nu \geq 3$), and $t_{\nu}$ is the univariate Student-$t$ distribution with degrees of freedom $\nu$. 

Both the Gaussian and the Student-$t$ copula functions belong to the elliptical family of copulas.
\subsection{Archimedean copulas}
Archimedean copulas are defined as follows,
\begin{equation}
C(\mathbf{u}|\phi) = \Psi^{-1}\left(\sum_{l=1}^L \Psi(u_l)\right),
\end{equation}
where we refer to $\Psi(\cdot)$ as the generator function and $\phi$ as the parameter of the copula. Some Archimedean copula functions are indicated in Table \ref{table:1} \cite{he2015heterogeneous}.
\begin{table*}[thb]
\caption{Archimedean copula functions.} 
\label{table:1}
	\centering
		\begin{tabular}{|c | c | c |}
		\hline

     Copula & Generator Function $\Psi$ &  Copulas in the Parametric Form \\
		\hline
 Clayton &   $\frac{1}{\phi} \left(u^{-\phi}-1 \right)$ &   $\left(\sum_{l=1}^L u_l^{-\phi} -1 \right)^{-\frac{1}{\phi}}, \hspace{2mm} \phi \in [-1,\infty) \backslash \{0\}$ \\[2ex]
 Frank   &  $-\log\frac{\exp\{-\phi u\}-1}{\exp\{-\phi\}-1}$  &
  $-\frac{1}{\phi}\log\left(1 + \frac{\prod_{l=1}^L{[\exp\{-\phi u_l\}-1]}}{\exp\{-\phi\}-1} \right), \phi \in \mathbb{R}\backslash \{0\}$ \\[2ex]
 Gumbel & $ -\ln u^{\phi}$ & $\exp\left\{- \left(\sum_{l=1}^L (-\ln u_l)^{\phi}\right)^{\frac{1}{\phi}}   \right\}, \hspace{2mm} \phi \in [1,\infty)$ \\[2ex]
 Independent & $-\ln u$ & $\prod_{l=1}^L u_l$ \\[2ex]
\hline
\end{tabular}
\end{table*}

\section{$A_{\mathbf{u}^t}$ in Log Test Statictics \eqref{eq:logT}}
\label{appendix:A}
First, we define $\tilde{\mathnormal{I}} = \{ ln | u_{ln} = 1, l = 1, 2, \ldots, L, n = 1, 2, \ldots, N \}$ as the index set with decisions $1$. Note that $\tilde{\mathnormal{I}} \cup \tilde{\mathnormal{I}}^c = \mathnormal{I}$. Moreover, let $\tilde{\mathnormal{I}}_j$ be the subset of $\tilde{\mathnormal{I}}$ and the cardinality of the set $\tilde{\mathnormal{I}}_j$ is $j$. We further define $\tilde{\mathnormal{I}}_{jt} = \{i_1n, i_2n, \ldots, i_tn, 0 \leq t \leq j \}$ as the subset of $\tilde{\mathnormal{I}}_j$ where $t = \left | \tilde{\mathnormal{I}}_{jt} \right |$. If $t$ is a even number, $\tilde{\mathnormal{I}}_{jt} = \tilde{\mathnormal{I}}_{jt}^e$, otherwise, $\tilde{\mathnormal{I}}_{jt} = \tilde{\mathnormal{I}}_{jt}^o$. Under hypothesis $H_1$, let $P_{\tilde{\mathnormal{I}}_{jt}(i_1n, i_2n, \ldots, i_tn)}$ denote the PMFs that only the decisions of $(i_1n, i_2n, \ldots, i_tn)$th sensors are $1$'s and that of the rest of sensors are $0$'s. Note $t = 0$ implies that no sensor makes a decision $1$ and $P_{\tilde{\mathnormal{I}}_{j0}}$ is used to denote all $0$ sensor decisions. Similarly, let $Q_{\tilde{\mathnormal{I}}_{jt}(i_1n, i_2n, \ldots, i_tn)}$ denote the PMFs under $H_0$.

In the following, we illustrate the process of obtaining $A_{\mathbf{u}^t} = \text{log}\frac{\prod_{0 \leq k \leq t}\mathcal{P}_{\tilde{\mathnormal{I}}_{tk}^e}^{(-1)^{t}} \prod_{0 \leq k \leq t}\mathcal{Q}_{\tilde{\mathnormal{I}}_{tk}^o}^{(-1)^{t}}}{\prod_{0 \leq k \leq t}\mathcal{Q}_{\tilde{\mathnormal{I}}_{tk}^e}^{(-1)^{t}} \prod_{0 \leq k \leq t}\mathcal{P}_{\tilde{\mathnormal{I}}_{tk}^o}^{(-1)^{t}}}$, where $\mathcal{P}_{\tilde{\mathnormal{I}}_{tk}} = \underset{\{ i_1n, i_2n, \ldots, i_kn \} \in \tilde{\mathnormal{I}}_{t}}{\prod}P_{\tilde{\mathnormal{I}}_{tk}(i_1n, i_2n, \ldots, i_kn)}$ and $\mathcal{Q}_{\tilde{\mathnormal{I}}_{tk}} = \underset{\{ i_1n, i_2n, \ldots, i_kn \} \in \tilde{\mathnormal{I}}_{t}}{\prod}Q_{\tilde{\mathnormal{I}}_{tk}(i_1n, i_2n, \ldots, i_kn)}$.

First, for $t = 1$, we have $k = 0, 1$. $A_{\mathbf{u}^1}$ is given as $A_{\mathbf{u}^1} = \text{log}\frac{Q_{\tilde{\mathnormal{I}}_{10}} P_{\tilde{\mathnormal{I}}_{11}(i_1n)}}{P_{\tilde{\mathnormal{I}}_{10}}Q_{\tilde{\mathnormal{I}}_{11}(i_1n)}}, 1 \leq i_1 \leq L$ which satisfies the $A_{\mathbf{u}^t}$ with $t = 1$.

For $t = 2$, we have $k = 0, 1, 2$. $A_{\mathbf{u}^2}$ is given as
$A_ {\mathbf{u}^2} = \text{log}\frac{P_{\tilde{\mathnormal{I}}_{20}}P_{\tilde{\mathnormal{I}}_{22}(i_1n, i_2n)}Q_{\tilde{\mathnormal{I}}_{21}(i_1n)}Q_{\tilde{\mathnormal{I}}_{21}(i_2n)}}{Q_{\tilde{\mathnormal{I}}_{20}}Q_{\tilde{\mathnormal{I}}_{22}({i_1n, i_2n})}P_{\tilde{\mathnormal{I}}_{21}(i_1n)}P_{\tilde{\mathnormal{I}}_{21}(i_2n)}}, 1 \leq i_1, i_2 \leq L$ which satisfies $A_{\mathbf{u}^t}$ with $t = 2$.

For $t = 3$, we have $k = 0, 1, 2, 3$. $A_{\mathbf{u}^3}$ is given as $A_{\mathbf{u}^3} = \text{log}\frac{\mathcal{Q}_{\tilde{\mathnormal{I}}^e_{30}}\mathcal{Q}_{\tilde{\mathnormal{I}}^e_{32}}\mathcal{P}_{\tilde{\mathnormal{I}}^o_{31}}\mathcal{P}_{\tilde{\mathnormal{I}}^o_{33}}}{\mathcal{P}_{\tilde{\mathnormal{I}}^e_{30}}\mathcal{P}_{\tilde{\mathnormal{I}}^e_{32}}\mathcal{Q}_{\tilde{\mathnormal{I}}^o_{31}}\mathcal{Q}_{\tilde{\mathnormal{I}}^o_{33}}}, 1 \leq i_1, i_2, i_3 \leq L$, where for the numerator, $\mathcal{Q}_{\tilde{\mathnormal{I}}^e_{30}} = Q_{\tilde{\mathnormal{I}}_{30}}$, $\mathcal{Q}_{\tilde{\mathnormal{I}}^e_{32}} = Q_{\tilde{\mathnormal{I}}_{32}(i_1n, i_2n)}Q_{\tilde{\mathnormal{I}}_{32}(i_1n, i_3n)}Q_{\tilde{\mathnormal{I}}_{32}(i_2n, i_3n)}$, $\mathcal{P}_{\tilde{\mathnormal{I}}^o_{31}} = P_{\tilde{\mathnormal{I}}_{31}(i_1n)}P_{\tilde{\mathnormal{I}}_{31}(i_2n)}P_{\tilde{\mathnormal{I}}_{31}(i_3n)}$ and $\mathcal{P}_{\tilde{\mathnormal{I}}^o_{33}} = P_{\tilde{\mathnormal{I}}_{33}(i_1n, i_2n, i_3n)}$. We can verify that $A_{\mathbf{u}^3}$ satisfies $A_{\mathbf{u}^t}$ with $t = 3$.


For $t = 4$, we have $k = 0, 1, 2, 3, 4$. $A_{\mathbf{u}^4}$ is given as 
$A_{\mathbf{u}^4} = \text{log}\frac{\mathcal{P}_{\tilde{\mathnormal{I}}^e_{40}}\mathcal{P}_{\tilde{\mathnormal{I}}^e_{42}}\mathcal{P}_{\tilde{\mathnormal{I}}^e_{44}}\mathcal{Q}_{\tilde{\mathnormal{I}}^o_{41}}\mathcal{Q}_{\tilde{\mathnormal{I}}^o_{43}}}{\mathcal{Q}_{\tilde{\mathnormal{I}}^e_{40}}\mathcal{Q}_{\tilde{\mathnormal{I}}^e_{42}}\mathcal{Q}_{\tilde{\mathnormal{I}}^e_{44}}\mathcal{P}_{\tilde{\mathnormal{I}}^o_{41}}\mathcal{P}_{\tilde{\mathnormal{I}}^o_{43}}}, 1 \leq i_1, i_2, i_3, i_4 \leq L$ which satisfies $A_{\mathbf{u}^t}$ with $t = 4$.


For $t = 5, 6, \ldots, L$, $A_{\mathbf{u}^t}$ can be easily verified.


\section{Proof of Theorem \ref{theorem:testS}}
\label{appendix:testS}
Note that \eqref{eq:logT} can be written as 
\begin{align}
\label{eq:logT2}
\text{log}\Lambda(\mathbf{u}) = \sum_{n = 1}^{N} \mathnormal{U}_n,
\end{align}
where $\mathnormal{U}_n = \sum_{\{ i_1n \} \in \mathnormal{I}_1 }A_{\mathbf{u}^1}\mathbf{u}^1 + \sum_{\{i_1n, i_2n\} \in \mathnormal{I}_2 }A_{\mathbf{u}^2}\mathbf{u}^2 +\ldots + \sum_{\{i_1n, i_2n, \ldots, i_tn\} \in \mathnormal{I}_t }A_{\mathbf{u}^t}\mathbf{u}^t + \ldots + \sum_{\{i_1n, i_2n, \ldots, i_Ln\} \in \mathnormal{I}_L }A_{\mathbf{u}^L}\mathbf{u}^L$.

\vspace{2mm}
Due to the assumption of temporal independence of sensor decisions, $\mathnormal{U}_n$ for all $1 \leq n \leq N$ are i.i.d. random variables. Hence, by applying the central limit theorem (CLT) \cite{papoulis2002probability}, $\text{log}\Lambda(\mathbf{u})$ is asymptotically Gaussian.

Note that $u_{ln}, l = 1, 2, \ldots, L$ are Bernoulli distributed under both hypotheses and can take a value of either $0$ or $1$ with certain probabilities. For the simplification of notation, we omit the time index $n$ here. For sensor decisions $s \in S$, we define $E = \{ j_1, j_2, \ldots, j_d, 1 \leq d \leq L \}$ as the index set when the sensor decisions of $s$ are $1$. Under $H_1$ hypothesis, the random variable $\mathnormal{U}_s = \sum_{\{ j_{m_1} \} \subset E }A_{\mathbf{u}^1} + \sum_{\{j_{m_1}, j_{m_2}\} \subset E }A_{\mathbf{u}^2} + \ldots + \sum_{\{j_{m_1}, j_{m_2}, \ldots, j_{m_{d-1}}\} \subset E }A_{\mathbf{u}^{d-1}} + \sum_{\{j_{m_1}, j_{m_2}, \ldots, j_{m_{d}}\} \subset E }A_{\mathbf{u}^{d}}$ with probability $P_s$ for $1 \leq d \leq L$, otherwise, $\mathnormal{U} = 0$ for $d = 0$. Similarly, we can obtain the values of $\mathnormal{U}$ under $H_0$ hypothesis. Since we can obtain the joint PMF of sensor decisions by integrating the joint PDF of their observations under both hypotheses, we now can evaluate the mean and variance of the Gaussian distributed fusion statistic under either hypothesis. The mean and variance of the fusion rule statistic under both hypotheses are given as follows
\begin{equation*}
\begin{aligned}
& \mu_0 = N\left [\sum_{s \in S }\mathnormal{U}_sQ_s\right ], \\
&\sigma_0^2 = N\left [\sum_{s \in S }\mathnormal{U}_s^2Q_s - (\mu_0/N)^2 \right], \\
& \mu_1 = N\left [\sum_{s \in S }\mathnormal{U}_sP_s \right ], \\
&\sigma_1^2 = N\left [\sum_{s \in S }\mathnormal{U}_s^2P_s - (\mu_1/N)^2 \right]. 
\end{aligned}
\end{equation*}

\bibliographystyle{IEEEbib}
\bibliography{refcopulaJ}

\begin{thebibliography}{10}

\bibitem{tenney1981detection}
Robert~R Tenney and Nils~R Sandell,
\newblock ``Detection with distributed sensors,''
\newblock {\em IEEE Transactions on Aerospace and Electronic systems}, , no. 4,
  pp. 501--510, 1981.

\bibitem{chair1986optimal}
Z~Chair and PK~Varshney,
\newblock ``Optimal data fusion in multiple sensor detection systems,''
\newblock {\em IEEE Transactions on Aerospace and Electronic Systems}, , no. 1,
  pp. 98--101, 1986.

\bibitem{hoballah1986neyman}
Imad~Y Hoballah and PK~Varshney,
\newblock ``Neyman-pearson detection wirh distributed sensors,''
\newblock in {\em Decision and Control, 1986 25th IEEE Conference on}. IEEE,
  1986, vol.~25, pp. 237--241.

\bibitem{viswanathan1997distributed}
Ramanarayanan Viswanathan and Pramod~K Varshney,
\newblock ``Distributed detection with multiple sensors part i. fundamentals,''
\newblock {\em Proceedings of the IEEE}, vol. 85, no. 1, pp. 54--63, 1997.

\bibitem{varshney2012distributed}
Pramod~K Varshney,
\newblock {\em Distributed detection and data fusion},
\newblock Springer Science \& Business Media, 2012.

\bibitem{chen2005optimality}
Biao Chen and Peter~K Willett,
\newblock ``On the optimality of the likelihood-ratio test for local sensor
  decision rules in the presence of nonideal channels,''
\newblock {\em IEEE Transactions on Information Theory}, vol. 51, no. 2, pp.
  693--699, 2005.

\bibitem{niu2005distributed}
Ruixin Niu and Pramod~K Varshney,
\newblock ``Distributed detection and fusion in a large wireless sensor network
  of random size,''
\newblock {\em EURASIP Journal on Wireless Communications and Networking}, vol.
  2005, no. 4, pp. 462--472, 2005.

\bibitem{niu2006distributed}
Ruixin Niu, Pramod~K Varshney, and Qi~Cheng,
\newblock ``Distributed detection in a large wireless sensor network,''
\newblock {\em Information Fusion}, vol. 7, no. 4, pp. 380--394, 2006.

\bibitem{chamberland2004asymptotic}
J-F Chamberland and Venugopal~V Veeravalli,
\newblock ``Asymptotic results for decentralized detection in power constrained
  wireless sensor networks,''
\newblock {\em IEEE Journal on selected areas in communications}, vol. 22, no.
  6, pp. 1007--1015, 2004.

\bibitem{yang2010distributed}
Yang Yang, Rick~S Blum, and Brian~M Sadler,
\newblock ``A distributed and energy-efficient framework for neyman-pearson
  detection of fluctuating signals in large-scale sensor networks,''
\newblock {\em IEEE Journal on Selected Areas in Communications}, vol. 28, no.
  7, 2010.

\bibitem{fang2012optimal}
Jun Fang, Hongbin Li, Zhi Chen, and Shaoqian Li,
\newblock ``Optimal precoding design and power allocation for decentralized
  detection of deterministic signals,''
\newblock {\em IEEE Transactions on Signal Processing}, vol. 60, no. 6, pp.
  3149--3163, 2012.

\bibitem{drakopoulos1991optimum}
ELIAS Drakopoulos and C-C Lee,
\newblock ``Optimum multisensor fusion of correlated local decisions,''
\newblock {\em IEEE Transactions on Aerospace and Electronic Systems}, vol. 27,
  no. 4, pp. 593--606, 1991.

\bibitem{kam1992optimal}
Moshe Kam, Qiang Zhu, and W~Steven Gray,
\newblock ``Optimal data fusion of correlated local decisions in multiple
  sensor detection systems,''
\newblock {\em IEEE Transactions on Aerospace and Electronic Systems}, vol. 28,
  no. 3, pp. 916--920, 1992.

\bibitem{willett2000good}
Peter Willett, Peter~F Swaszek, and Rick~S Blum,
\newblock ``The good, bad and ugly: distributed detection of a known signal in
  dependent gaussian noise,''
\newblock {\em IEEE Transactions on Signal Processing}, vol. 48, no. 12, pp.
  3266--3279, 2000.

\bibitem{yan2001distributed}
Qing Yan and Rick~S Blum,
\newblock ``Distributed signal detection under the neyman-pearson criterion,''
\newblock {\em IEEE Transactions on Information Theory}, vol. 47, no. 4, pp.
  1368--1377, 2001.

\bibitem{chamberland2006dense}
J-F Chamberland and Venugopal~V Veeravalli,
\newblock ``How dense should a sensor network be for detection with correlated
  observations?,''
\newblock {\em IEEE Transactions on Information Theory}, vol. 52, no. 11, pp.
  5099--5106, 2006.

\bibitem{khalid2012cooperative}
Lamiaa Khalid and Alagan Anpalagan,
\newblock ``Cooperative sensing with correlated local decisions in cognitive
  radio networks,''
\newblock {\em IEEE Transactions on Vehicular Technology}, vol. 61, no. 2, pp.
  843--849, 2012.

\bibitem{kasasbeh2016hard}
Hadi Kasasbeh, Lei Cao, and Ramanarayanan Viswanathan,
\newblock ``Hard decision based distributed detection in multi-sensor system
  over noise correlated sensing channels,''
\newblock in {\em Information Science and Systems (CISS), 2016 Annual
  Conference on}. IEEE, 2016, pp. 280--285.

\bibitem{kasasbeh2017soft}
Hadi Kasasbeh, Lei Cao, and Ramanarayanan Viswanathan,
\newblock ``Soft-decision based distributed detection over correlated sensing
  channels,''
\newblock in {\em Information Sciences and Systems (CISS), 2017 51st Annual
  Conference on}. IEEE, 2017, pp. 1--6.

\bibitem{veeravalli2012distributed}
Venugopal~V Veeravalli and Pramod~K Varshney,
\newblock ``Distributed inference in wireless sensor networks,''
\newblock {\em Phil. Trans. R. Soc. A}, vol. 370, no. 1958, pp. 100--117, 2012.

\bibitem{tsitsiklis1985complexity}
John Tsitsiklis and Michael Athans,
\newblock ``On the complexity of decentralized decision making and detection
  problems,''
\newblock {\em IEEE Transactions on Automatic Control}, vol. 30, no. 5, pp.
  440--446, 1985.

\bibitem{nelsen2013introduction}
Roger~B Nelsen,
\newblock {\em An introduction to copulas}, vol. 139,
\newblock Springer Science \& Business Media, 2013.

\bibitem{iyengar2011parametric}
Satish~G Iyengar, Pramod~K Varshney, and Thyagaraju Damarla,
\newblock ``A parametric copula-based framework for hypothesis testing using
  heterogeneous data,''
\newblock {\em IEEE Transactions on Signal Processing}, vol. 59, no. 5, pp.
  2308--2319, 2011.

\bibitem{he2012fusing}
Hao He, Arun Subramanian, Pramod~K Varshney, and Thyagaraju Damarla,
\newblock ``Fusing heterogeneous data for detection under non-stationary
  dependence,''
\newblock in {\em 2012 15th International Conference on Information Fusion
  (FUSION)}. IEEE, 2012, pp. 1792--1799.

\bibitem{sundaresan2011location}
Ashok Sundaresan and Pramod~K Varshney,
\newblock ``Location estimation of a random signal source based on correlated
  sensor observations,''
\newblock {\em IEEE Transactions on Signal Processing}, vol. 59, no. 2, pp.
  787--799, 2011.

\bibitem{iyengar2011biometric}
Satish~G Iyengar, Pramod~K Varshney, and Thyagaraju Damarla,
\newblock ``Biometric authentication: a copula based approach,''
\newblock {\em Multibiometrics for Human Identification}, pp. 95--119, 2011.

\bibitem{he2015social}
Hao He, Arun Subramanian, Sora Choi, Pramod~K Varshney, and Thyagaraju Damarla,
\newblock ``Social media data assisted inference with application to stock
  prediction,''
\newblock in {\em 2015 49th Asilomar Conference on Signals, Systems and
  Computers}. IEEE, 2015, pp. 1801--1805.

\bibitem{bedford2001probability}
Tim Bedford and Roger~M Cooke,
\newblock ``Probability density decomposition for conditionally dependent
  random variables modeled by vines,''
\newblock {\em Annals of Mathematics and Artificial intelligence}, vol. 32, no.
  1-4, pp. 245--268, 2001.

\bibitem{bedford2002vines}
Tim Bedford and Roger~M Cooke,
\newblock ``Vines: A new graphical model for dependent random variables,''
\newblock {\em Annals of Statistics}, pp. 1031--1068, 2002.

\bibitem{aas2009pair}
Kjersti Aas, Claudia Czado, Arnoldo Frigessi, and Henrik Bakken,
\newblock ``Pair-copula constructions of multiple dependence,''
\newblock {\em Insurance: Mathematics and economics}, vol. 44, no. 2, pp.
  182--198, 2009.

\bibitem{subramanian2011fusion}
Arun Subramanian, Ashok Sundaresan, and Pramod~K Varshney,
\newblock ``Fusion for the detection of dependent signals using multivariate
  copulas,''
\newblock in {\em 2011 Proceedings of the 14th International Conference on
  Information Fusion (FUSION)}. IEEE, 2011, pp. 1--8.

\bibitem{sundaresan2011copula}
Ashok Sundaresan, Pramod~K Varshney, and Nageswara~SV Rao,
\newblock ``Copula-based fusion of correlated decisions,''
\newblock {\em IEEE Transactions on Aerospace and Electronic Systems}, vol. 47,
  no. 1, pp. 454--471, 2011.

\bibitem{he2015heterogeneous}
Hao He,
\newblock {\em Heterogeneous sensor signal processing for inference with
  nonlinear dependence},
\newblock Ph.D. thesis, Syracuse University, 2015.

\bibitem{dissmann2013selecting}
Jeffrey Dissmann, Eike~C Brechmann, Claudia Czado, and Dorota Kurowicka,
\newblock ``Selecting and estimating regular vine copulae and application to
  financial returns,''
\newblock {\em Computational Statistics \& Data Analysis}, vol. 59, pp. 52--69,
  2013.

\bibitem{mari2001correlation}
Dominique~Drouet Mari and Samuel Kotz,
\newblock {\em Correlation and dependence}, vol. 518,
\newblock World Scientific, 2001.

\bibitem{joe1996families}
Harry Joe,
\newblock ``Families of m-variate distributions with given margins and m
  (m-1)/2 bivariate dependence parameters,''
\newblock {\em Lecture Notes-Monograph Series}, pp. 120--141, 1996.

\bibitem{morales2010bayesian}
Oswaldo Morales~N{\'a}poles,
\newblock {\em Bayesian belief nets and vines in aviation safety and other
  applications},
\newblock TU Delft, Delft University of Technology, 2010.

\bibitem{varshney1997distributed}
Pramod~K Varshney,
\newblock ``Distributed bayesian detection: Parallel fusion network,''
\newblock in {\em Distributed Detection and Data Fusion}, pp. 36--118.
  Springer, 1997.

\bibitem{morales2010number}
Oswaldo Morales~Napoles, Roger~M Cooke, and Dorota Kurowicka,
\newblock ``About the number of vines and regular vines on n nodes,''
\newblock 2010.

\bibitem{akaike1973information}
Hirotogu Akaike, BN~Petrov, and F~Csaki,
\newblock ``Information theory and an extension of the maximum likelihood
  principle,''
\newblock 1973.

\bibitem{czado2013selection}
Claudia Czado, Eike~Christian Brechmann, and Lutz Gruber,
\newblock ``Selection of vine copulas,''
\newblock in {\em Copulae in Mathematical and Quantitative Finance}, pp.
  17--37. Springer, 2013.

\bibitem{czado2013selection2}
Claudia Czado, Stephan Jeske, and Mathias Hofmann,
\newblock ``Selection strategies for regular vine copulae,''
\newblock {\em Journal de la Soci{\'e}t{\'e} Fran{\c{c}}aise de Statistique},
  vol. 154, no. 1, pp. 174--191, 2013.

\bibitem{wassermann2006all}
L~Wassermann,
\newblock ``All of nonparametric statistics,'' 2006.

\bibitem{schwarz1978estimating}
Gideon Schwarz et~al.,
\newblock ``Estimating the dimension of a model,''
\newblock {\em The annals of statistics}, vol. 6, no. 2, pp. 461--464, 1978.

\bibitem{flach2005repairing}
Peter~A Flach and Shaomin Wu,
\newblock ``Repairing concavities in roc curves.,''
\newblock in {\em IJCAI}, 2005, pp. 702--707.

\bibitem{fawcett2006introduction}
Tom Fawcett,
\newblock ``An introduction to roc analysis,''
\newblock {\em Pattern recognition letters}, vol. 27, no. 8, pp. 861--874,
  2006.

\bibitem{papoulis2002probability}
Athanasios Papoulis and S~Unnikrishna Pillai,
\newblock {\em Probability, random variables, and stochastic processes},
\newblock Tata McGraw-Hill Education, 2002.

\end{thebibliography}

\end{document}